%% file: main.tex
\documentclass[10pt,journal,compsoc]{IEEEtran}

\usepackage{cite}
\usepackage{makecell}
\usepackage{url}
\usepackage{graphicx}
\usepackage{multirow,multicol}
\renewcommand\arraystretch{1.8}
\usepackage{amssymb}
\usepackage{amsmath,amsfonts,amsthm,mathtools}
\usepackage{utfsym}
\usepackage{float}
\usepackage{booktabs}
\usepackage{enumitem} 
\setlist[enumerate]{leftmargin=*} 
\setlist[itemize]{leftmargin=*} 

\setlist[itemize]{leftmargin=*,itemsep=2pt,topsep=2pt,parsep=0pt}
\setlist[enumerate]{leftmargin=*,itemsep=2pt,topsep=2pt,parsep=0pt}
\begin{document}
\title{Service Ecosystem Evolution: A Comprehensive Survey from Complex Network Perspectives}

\author{Shuiguang~Deng,~\IEEEmembership{Senior Member,~IEEE,}
        Xingliang~Wang,
        Guochang~Li,
        Yunkun~Wang,
        Haoran~Xu,
        Chen~Zhi,
        Junxiao~Han
        and~Jianwei~Yin

\IEEEcompsocitemizethanks {
  \IEEEcompsocthanksitem S. Deng is with the Hainan Institute of Zhejiang University, Sanya, China, and the College of Computer Science and Technology, Zhejiang University, Hangzhou, China. X. Wang, G. Li, Y. Wang, H. Xu, C. Zhi, and J. Yin are with the College of Computer Science and Technology, Zhejiang University, Hangzhou, China.\protect \quad E-mail: \{dengsg, wangxingliang, gcli, wangykun, haoran.x, zjuzhichen, zjuyjw\}@zju.edu.cn. J. Han is with the School of Computer \& Computing Science, Hangzhou City University, Hangzhou 310015, China.\protect \quad E-mail: hanjx@hzcu.edu.cn.
  (Corresponding authors: Chen Zhi and Junxiao Han.)
}
}


\IEEEtitleabstractindextext{%
\begin{abstract}

Digital society increasingly relies on complex service ecosystems, formed by interconnected services from technology giants. However, the growing scale and intricate dependencies of these ecosystems pose significant challenges to their evolution, frequently leading to systemic failures during upgrades or restructuring. To address these challenges, current research is shifting from the perspective of single services to the ecosystem. 
Drawing upon the synthesis and induction of current research, we present a comprehensive survey focused on the study of service ecosystem evolution, employing a novel three-stage analytical framework. This framework structures the evolutionary lifecycle and provides a systematic way to organize and review existing research, filling the gap caused by the lack of surveys specifically focused on service ecosystem evolution.
Additionally, we pioneer the application of complex network theory to analyze service ecosystem evolution, providing a novel perspective to capture the inherent connectivity, dynamics, and emergent properties often missed by traditional approaches. 
Building on our analysis, we point out critical research gaps and propose three specific future directions. We provide a robust theoretical foundation and methodological guidance for understanding and guiding the sustainable evolution of modern service ecosystems.

\end{abstract}

}

\maketitle

\IEEEdisplaynontitleabstractindextext

\IEEEpeerreviewmaketitle

\input{section/sec_introduction.tex}
\input{section/sec_framework.tex}

\input{section/sec_representation.tex}

\input{section/sec_perception.tex}
\input{section/sec_implementation.tex}
\input{section/sec_evaluation.tex}
\input{section/sec_opportunities.tex}

\input{section/sec_conclusion.tex}

\section*{Acknowledgments}
This work was supported in part by the National Natural Science Foundation of China under Grant 62125206.

\bibliographystyle{IEEEtran}
{\footnotesize
\setlength{\itemsep}{0pt plus 0.3ex}
\bibliography{references}}


\begin{IEEEbiography}
[{\includegraphics[width=0.9in,height=1.1in,clip,keepaspectratio]{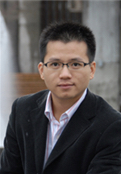}}]{Shuiguang Deng}
	(Senior Member, IEEE) received the B.S. and Ph.D. degrees in computer science from the College of Computer Science and Technology, Zhejiang University, China, in 2002 and 2007, respectively. He is currently affiliated with the Hainan Institute of Zhejiang University, Sanya, China, and is a full professor at the College of Computer Science and Technology, Zhejiang University, China. He was a visiting scholar at the Massachusetts Institute of Technology, Cambridge, MA, USA, in 2014 and at Stanford University, Stanford, CA, USA, in 2015. His research interests include edge computing, service computing, cloud computing, and business process management. He serves as an associate editor for IEEE Transactions on Services Computing, Knowledge and Information Systems, Computing, and IET Cyber-Physical Systems: Theory \& Applications. To date, he has published more than 100 papers in journals and refereed conference proceedings. He received the Rising Star Award from the IEEE Technical Community on Services Computing in 2018. He is a Fellow of the IET.
\end{IEEEbiography}

\vspace{-4mm}

\begin{IEEEbiography}[{\includegraphics[width=0.9in,height=1.1in,clip,keepaspectratio]{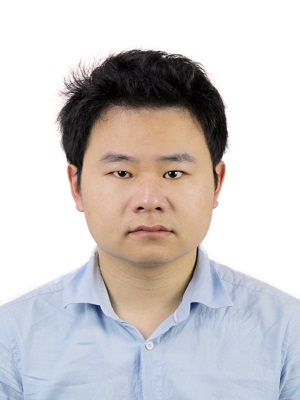}}]{Xingliang Wang}
	received the B.S. and M.S. degrees from Shandong University of Technology and Hangzhou Dianzi University, China, in 2019 and 2022, respectively. He is currently pursuing the Ph.D. degree at Zhejiang University. His research interests include graph neural networks, service computing, and interpretable machine learning.
\end{IEEEbiography}

\vspace{-4mm}

\begin{IEEEbiography}[{\includegraphics[width=0.9in,height=1.1in,clip,keepaspectratio]{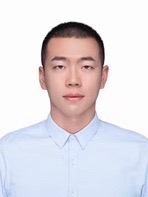}}]{Guochang Li}
    received the B.S. degree from Zhejiang University, China, in 2021. He is currently pursuing the Ph.D. degree at Zhejiang University. His research interests include automated program repair and agent-based software engineering.
\end{IEEEbiography}

\vspace{-4mm}

\begin{IEEEbiography}[{\includegraphics[width=0.9in,height=1.1in,clip,keepaspectratio]{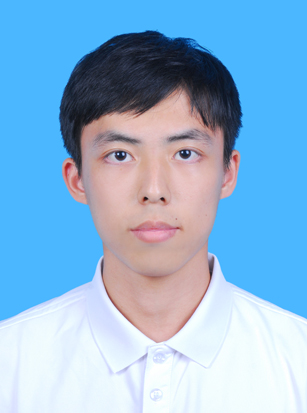}}]{Yunkun Wang}
	 received the B.S. degree from Beijing University of Posts and Telecommunications, China, in 2022. He is currently pursuing the Ph.D. degree at Zhejiang University. His research interests include software engineering and artificial intelligence.
\end{IEEEbiography}

\vspace{-4mm}

\begin{IEEEbiography}[{\includegraphics[width=0.9in,height=1.1in,clip,keepaspectratio]{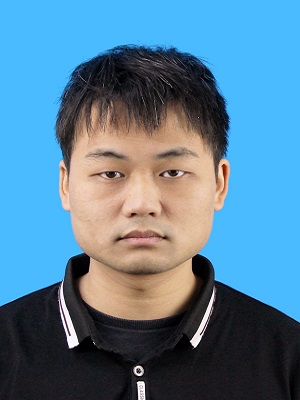}}]{Haoran Xu}
	 received the B.S. and M.S. degrees from Hangzhou Dianzi University, China, in 2019 and 2022, respectively. He is currently pursuing the Ph.D. degree at Zhejiang University. His research interests include software engineering and recommender systems.
\end{IEEEbiography}

\vspace{-4mm}

\begin{IEEEbiography}[{\includegraphics[width=0.9in,height=1.1in,clip,keepaspectratio]{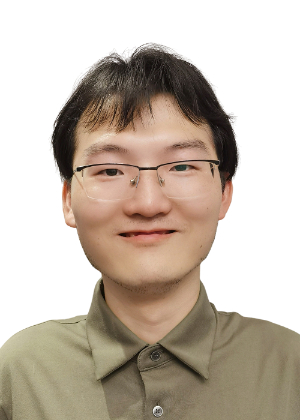}}]{Chen Zhi} received the Ph.D. degree in computer science from Zhejiang University in 2022. He is currently an associate research fellow at the School of Software Technology, Zhejiang University. He has published more than 10 papers in international journals and conference proceedings. His research interests include service computing, software testing, and program analysis.
\end{IEEEbiography}

\vspace{-4mm}

\begin{IEEEbiography}[{\includegraphics[width=0.9in,height=1.1in,clip,keepaspectratio]{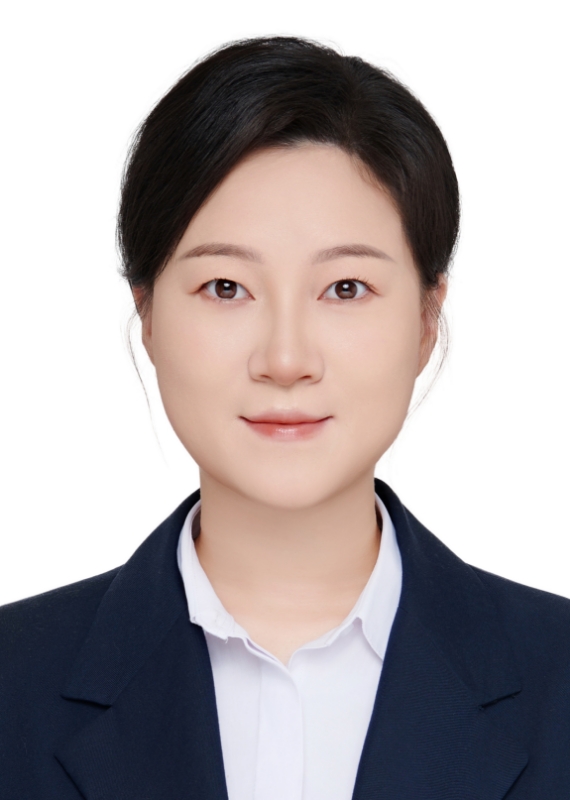}}]{Junxiao Han}
is currently a lecturer at the School of Computer \& Computing Science, Hangzhou City University, Hangzhou, China. She received the Ph.D. degree from the College of Computer Science and Technology, Zhejiang University, Hangzhou, China, in 2021. Her research interests include AI for software engineering and mining software repositories. She serves as a reviewer for several international journals, including IEEE Transactions on Software Engineering, Empirical Software Engineering, and Automated Software Engineering.
\end{IEEEbiography}

\vspace{-4mm}

\begin{IEEEbiography}[{\includegraphics[width=0.9in,height=1.1in,clip,keepaspectratio]{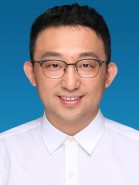}}]{Jianwei Yin}
	received the Ph.D. degree in computer science from Zhejiang University (ZJU), Hangzhou, China, in 2001. He was a visiting scholar at the Georgia Institute of Technology, Atlanta, GA, USA. He is currently a full professor at the College of Computer Science, ZJU. To date, he has published more than 100 papers in leading international journals and conference proceedings. His research interests include service computing and business process management.
    Prof. Yin is an Associate Editor of the IEEE TRANSACTIONS ON SERVICES COMPUTING.
\end{IEEEbiography}

\end{document}

%% file: section/sec_introduction.tex
\section{Introduction}
\label{sec:introduction}

With the rapid development of information technology, modern society increasingly relies on diverse services, and internet giants (e.g., Alibaba) keep expanding their service boundaries into multifaceted ecosystems covering e-commerce, cloud computing, finance, and entertainment.

However, the exponential growth in the number of services and their increasingly complex interdependencies pose severe challenges to the evolution of service ecosystems. Large-scale systemic failures have frequently occurred in internet giants~\cite{schroeder2009large,liu2021microhecl}, whose root causes can often be traced back to the service evolution process, such as inappropriate resource allocation, unresolved dependency conflicts, or inter-service interface incompatibilities, reflecting the difficulties in guiding and managing ecosystem evolution.

To address the aforementioned challenges, the research paradigm in service computing is gradually shifting from focusing on single service optimization~\cite{deb2013evolutionary,barcelo2016iot} or simple service composition~\cite{gao2017service,lu2020dcem} towards examining the structure, function, and evolutionary characteristics of services from a holistic ecosystem perspective~\cite{boley2007digital,zhong2014time,geary2020guide}.

However, the absence of a universally accepted definition for ``service ecosystem'' has led to fragmentation across different research perspectives~\cite{xue2022research,abeywickrama2017survey,li2011research}. Based on previous research, we define it as a complex adaptive system composed of various software services. This system comprises diverse heterogeneous entities (including service providers, consumers, brokers, infrastructure operators, etc.) and dynamic interaction relationships. Within this ecosystem, entities are interdependent and co-evolve, responding to continuously changing market demands and environmental pressures through the value co-creation mechanisms, together shaping an organic whole.


Based on the above definitions, we formally define a service ecosystem at time $t \in T$ as
\[
E(t)=\langle S,A,\rho,\Phi(t)\rangle,
\]
where $S=\{S_i\}_{i=1}^{n}$ is the set of services and each service is represented as $S_i=(I_i,O_i,F_i,X_i)$. Here, $I_i$ and $O_i$ denote the input and output sets, $F_i:I_i\rightarrow O_i$ is the functional mapping, and $X_i$ denotes the service attributes. The set $A=\{a_\ell\}_{\ell=1}^{m}$ contains the ecosystem actors, such as service providers, brokers, consumers, and infrastructure operators. Let $\mathcal{R}=\{\text{provider},\text{broker},\text{consumer},\text{infrastructure}\}$ be the set of basic roles; the mapping $\rho:A\rightarrow 2^{\mathcal{R}}$ assigns one or more roles to each actor. With $V=S\cup A$, the dynamic interaction matrix is defined as $\Phi(t)=[\phi_{uv}(t)]_{u,v\in V}$, where $\phi_{uv}(t)\in\mathbb{R}_{\geq 0}$ represents the presence and strength of the interaction from entity $u$ to entity $v$ at time $t$, with zero indicating no observed interaction.

Despite the attention garnered by service ecosystems, in-depth and systematic understanding of the drivers, mechanisms, and principles of their evolution remains lacking. While a few surveys have touched upon service ecosystems~\cite{xue2022research,golgeci2022bibliometric}, none concentrate on ``evolution'' as the core subject---yet this process is precisely where many current ecosystem challenges arise. Drawing upon the synthesis and induction of current research, we propose a three-stage analytical framework: $Evolution_{SE} = \mathcal{F}(Perception, Implementation, Evaluation)$, decomposing the evolutionary process into three interconnected phases---Evolution Perception, Evolution Implementation, and Evolution Evaluation---forming a dynamic closed-loop feedback system (detailed in Section~\ref{Service_Ecosystem_Evolution_Framework}).

Given that service ecosystems exhibit the structural characteristics of dynamically evolving complex networks, a systematic analysis is particularly crucial. We innovatively employ complex network theory~\cite{erdHos1960evolution,watts1998collective,barabasi1999emergence} as the core analytical perspective to review and classify existing research on service ecosystem evolution. By leveraging network analysis methods~\cite{zhou2020measurement,adeleye2021constructing}, we aim to integrate the literature from a complex network standpoint, revealing the focus and contributions of this research across different elements within the complex network framework. This network-based perspective is intended to provide subsequent researchers with a more specialized theoretical foundation and toolset, assisting them in grasping the open issues and future directions in evolution research from a network perspective.

The main contributions of this paper include:

\begin{itemize}
  
  \item \textbf{Comprehensive Literature Analysis}: We conduct a thorough and systematic review of existing research on service ecosystem evolution, encompassing key aspects such as identifying evolutionary drivers, selecting and executing evolutionary strategies, and evaluating the outcomes of evolution. We synthesize a broad range of studies, revealing specific differences and deficiencies in current research regarding method selection, applicable scenarios, and perspectives from complex networks, offering a holistic overview of this field.
  \item \textbf{Novel Three-Stage Analytical Framework}: We introduce a novel framework, denoted as $Evolution_{SE} = \mathcal{F}(Perception, Implementation, Evaluation)$. This framework structures our analysis by categorizing existing research according to the distinct stages of the evolutionary lifecycle, systematically integrating existing fragmented research, and enhancing the clarity and systematic nature of the review.
  \item \textbf{New Research Perspective via Complex Networks}: We introduce complex network theory for analyzing service ecosystem evolution. This perspective allows viewing the ecosystem as a multi-layered complex network, enabling a deeper understanding of its connectivity, dynamics, and emergent properties using complex network tools, moving beyond traditional isolated or static service paradigms and offering novel insights into the underlying mechanisms driving evolution.
  \item \textbf{Identification of Open Issues and Future Opportunities}: Based on our comprehensive analysis structured by the three-stage framework and informed by the complex network perspective, we identify critical research gaps, open issues, and promising future research directions. This contribution aims to guide and stimulate subsequent investigations in the field of service ecosystem evolution.

\end{itemize}

\textbf{Difference to Existing Surveys}: While prior surveys on service ecosystems~\cite{xue2022research,golgeci2022bibliometric} exist, none focus on evolution. This survey differs in three ways: (i) it proposes a three-stage ``Perception–Implementation–Evaluation'' framework that organizes research across the full evolutionary lifecycle; (ii) it adopts a complex-network perspective to classify existing work and examine their limitations in network representation and analysis; (iii) by combining both, it distinguishes itself in research theme, organizational structure, and analytical depth.

The remainder of this paper is organized as follows. Section~\ref{Service_Ecosystem_Evolution_Framework} introduces the proposed three-stage analytical framework that forms the organizational backbone of this survey. Section~\ref{sec:reprent} examines service ecosystem representation from a complex network perspective, categorizing current methods into static and dynamic network information-based approaches. Following the stages defined in our framework, Sections~\ref{sec:Evolution_Perception}, \ref{sec_Evolution_Implementation}, and \ref{sec_Evolution_Evaluation} provide a systematic review of evolution perception, implementation, and evaluation, respectively. Section~\ref{sec_Open_Issues_and_Opportunities} synthesizes our findings to identify critical research gaps and future opportunities. Finally, Section~\ref{sec:conclusion} concludes the paper.


%% file: section/sec_framework.tex
\section{Service Ecosystem Evolution Framework}
\label{Service_Ecosystem_Evolution_Framework}

\begin{figure*}[htbp]
  \centering
  \includegraphics[width=0.95\textwidth]{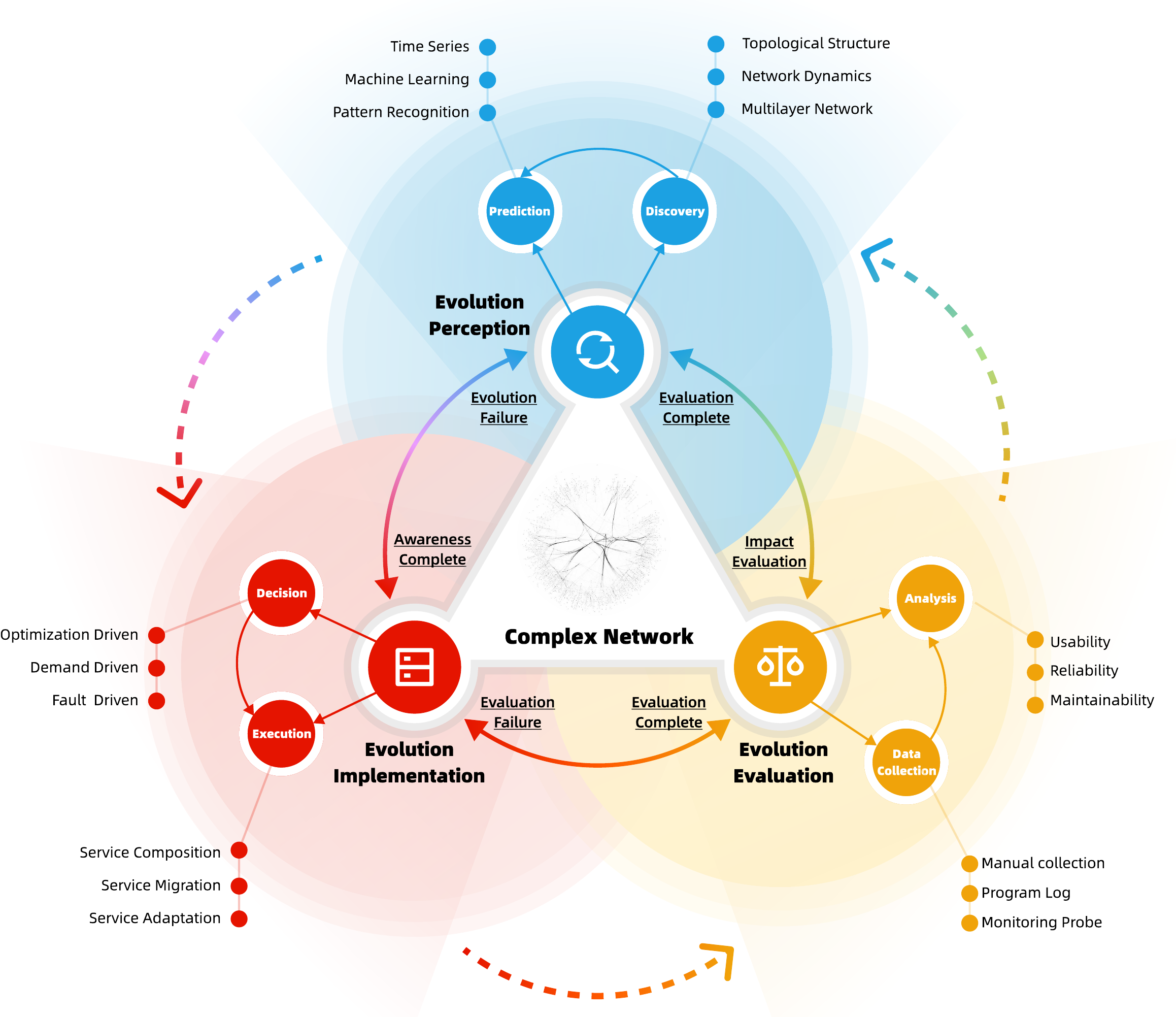}
  \caption{The framework divides the life cycle of service
  ecosystem evolution into three stages: (1) evolution
  perception, (2) evolution implementation, and (3) evolution
  evaluation.}
  \label{fig_framework}
\end{figure*}

To systematically organize the diverse research on service
ecosystem evolution, we propose a three-stage analytical
framework: $Evolution_{SE} = \mathcal{F}(Perception,
Implementation, Evaluation)$, as illustrated in
Figure~\ref{fig_framework}. This framework decomposes the
evolutionary lifecycle into three interconnected phases,
with complex network theory providing foundational
analytical support across all stages.

\begin{itemize}

\item \textbf{Evolution Perception} (Blue Area in
Figure~\ref{fig_framework}): This stage identifies and
predicts evolutionary dynamics and potential trends within
the service ecosystem~\cite{zhou2020evolutionary}, using
techniques such as network analysis, time series analysis,
and machine learning to detect signals of change and
provide inputs for subsequent decision-making.

\item \textbf{Evolution Implementation} (Red Area in
Figure~\ref{fig_framework}): This stage formulates and
executes specific evolution strategies. It integrates
perceived state information, external driving forces, and
operational constraints to determine evolutionary
actions~\cite{Papageorgiou2014DecisionSF}, and then executes
targeted activities such as service migration, composition,
or adaptation to modify the ecosystem.

\item \textbf{Evolution Evaluation} (Yellow Area in
Figure~\ref{fig_framework}): This stage measures the impact
of implemented changes by collecting post-evolution data
(e.g., logs, monitoring probes) and evaluating it against
predefined metrics~\cite{xue2020value,xue2020integrative}
(e.g., usability, reliability, maintainability) to ascertain
whether evolutionary objectives were met.

\end{itemize}

The typical evolution process follows the sequential flow
depicted by the outermost arrows in
Figure~\ref{fig_framework}: perceiving whether evolution is
occurring, formulating and executing strategies, and finally
assessing outcomes. Importantly, the three stages are not
strictly linear; they can iterate, forming a closed-loop
feedback system. For instance, should evaluation results
fail to meet predefined criteria, the process may loop back
to the implementation stage for re-execution, ensuring
continuous adaptation and sustainable evolution.

%% file: section/sec_representation.tex
\section{Service Ecosystem Representation}
\label{sec:reprent}
The service ecosystem exhibits complex characteristics, such as large scale and irregular topology. As the representation of the ecosystem fundamentally determines our ability to perceive, implement, and evaluate its evolution, establishing an effective representation is a crucial prerequisite for the three-stage evolutionary analysis proposed in this survey. Complex network analysis provides a framework capable of understanding the overall behavior and structural characteristics of the service ecosystem, which comprehensively describes interactions among services and the dynamic evolution of services, thereby becoming the preferred method.

Complex networks are mathematical models used to study systems composed of a large number of nodes and connections, while traditional graph theory primarily focuses on the theoretical study of small-scale static graphs. The random graph model~\cite{erdHos1960evolution} is one of the earliest well-known complex network models, and the most popular complex network models are the small-world network model~\cite{watts1998collective} and the scale-free network model~\cite{barabasi1999emergence}, which reveal the "small-world" phenomenon and scale-free properties of the complex network. 

Based on the temporal characteristics of the information leveraged, we categorize existing representation methods into two groups. Section~\ref{sec:static_rep} examines \textit{static network information-based representation}, which focuses on network topology and node attributes at a specific point in time. Section~\ref{sec:dynamic_rep} discusses \textit{dynamic network information-based representation}, which incorporates the temporal dimension to capture how network structures evolve over time. Table~\ref{table_evolution_representation} summarizes and compares selected studies. The abbreviations used in the table are as follows: UT (Unstructured Textual information from the service ecosystem), SC (Structural Characteristics of the service ecosystem), NP (Network Properties of the service ecosystem), Topic (Topic Model), CN (Complex Network model), and DG (Dynamic Model).

\begin{table*}[!htbp]
\centering
\caption{Summary of the work on service ecosystem representation}
\label{table_evolution_representation}
\footnotesize
\renewcommand{\arraystretch}{1.15}
\setlength{\tabcolsep}{0.8mm}{
\begin{tabular}{c|c|cccc|c|cccc|ccc}
    \toprule
    \multicolumn{2}{c|}{\multirow{2}*{Related Work}}
    & \multicolumn{4}{c|}{\multirow{2}*{Model name}} 
    & \multirow{2}*{Year}
    & \multicolumn{4}{c|}{Represented Properties}
    & \multicolumn{3}{c}{Methodology} 
    \\

    \multicolumn{2}{c|}{} 
    & \multicolumn{4}{c|}{} 
    & 
    & UT & SC & NP & Others
    & Topic & CN & DG
    \\ 
    \midrule

    \multirow{13}*{\makecell{Static \\ Network \\ Information \\Based}}
    & Cao et al.~\cite{cao2019qos}
    & \multicolumn{4}{c|}{Relational Topic Model} 
    & 2019
    & \checkmark & \scalebox{0.75}{\usym{2613}} & \scalebox{0.75}{\usym{2613}} & \scalebox{0.75}{\usym{2613}}
    & \checkmark& \scalebox{0.75}{\usym{2613}} & \scalebox{0.75}{\usym{2613}}
    \\   
    & Jiang et al.~\cite{jiang2019cloud}
    & \multicolumn{4}{c|}{Hierarchical Dirichlet Processes} 
    & 2019
    & \checkmark & \scalebox{0.75}{\usym{2613}} & \scalebox{0.75}{\usym{2613}} & \scalebox{0.75}{\usym{2613}}
    & \checkmark& \scalebox{0.75}{\usym{2613}} & \scalebox{0.75}{\usym{2613}}
    \\  
    & Yang et al.~\cite{yang2021web}
    & \multicolumn{4}{c|}{Biterm Topic Model} 
    & 2021
    & \checkmark & \scalebox{0.75}{\usym{2613}} & \scalebox{0.75}{\usym{2613}} & \scalebox{0.75}{\usym{2613}}
    & \checkmark& \scalebox{0.75}{\usym{2613}} & \scalebox{0.75}{\usym{2613}}
    \\  
    & Shen et al.~\cite{shen2022picf}
    & \multicolumn{4}{c|}{PICF-LDA} 
    & 2022
    & \checkmark &  \scalebox{0.75}{\usym{2613}} & \scalebox{0.75}{\usym{2613}} & \scalebox{0.75}{\usym{2613}}
    & \checkmark& \scalebox{0.75}{\usym{2613}} & \scalebox{0.75}{\usym{2613}}
    \\  
    & Zhang et al.~ \cite{zhang2020learning}
    & \multicolumn{4}{c|}{SR-LDA} 
    & 2020
    & \checkmark &  \scalebox{0.75}{\usym{2613}} & \scalebox{0.75}{\usym{2613}} & \scalebox{0.75}{\usym{2613}}
    & \checkmark& \scalebox{0.75}{\usym{2613}} & \scalebox{0.75}{\usym{2613}}
    \\  
    & Chen et al.~\cite{chen2013constructing}
    & \multicolumn{4}{c|}{Global Service Ecosystem} 
    & 2013
    & \scalebox{0.75}{\usym{2613}}& \scalebox{0.75}{\usym{2613}} & \checkmark & \scalebox{0.75}{\usym{2613}}
    & \scalebox{0.75}{\usym{2613}} & \checkmark & \scalebox{0.75}{\usym{2613}}
    \\  
    & Lyu et al.~\cite{lyu2014three}
    & \multicolumn{4}{c|}{Multi-layer Network} 
    & 2014
    & \scalebox{0.75}{\usym{2613}}& \checkmark & \scalebox{0.75}{\usym{2613}} & \scalebox{0.75}{\usym{2613}}
    & \scalebox{0.75}{\usym{2613}} & \checkmark & \scalebox{0.75}{\usym{2613}}
    \\  
    & Liu et al.~\cite{liu2023data}
    & \multicolumn{4}{c|}{Multi-layer Network} 
    & 2023
    & \scalebox{0.75}{\usym{2613}} & \checkmark & \scalebox{0.75}{\usym{2613}} & \scalebox{0.75}{\usym{2613}}
    & \scalebox{0.75}{\usym{2613}}& \checkmark  & \scalebox{0.75}{\usym{2613}}
    \\  
    & Adeleye et al.~\cite{adeleye2019fitness}
    & \multicolumn{4}{c|}{Bianconi-Barabási Complex Network Model} 
    & 2019
    & \scalebox{0.75}{\usym{2613}}& \scalebox{0.75}{\usym{2613}} & \checkmark & \scalebox{0.75}{\usym{2613}}
    & \scalebox{0.75}{\usym{2613}}& \checkmark  & \scalebox{0.75}{\usym{2613}}
    \\  
    & Adeleye et al.~\cite{adeleye2021constructing}
    & \multicolumn{4}{c|}{Bianconi-Barabási Complex Network Model} 
    & 2021
    & \scalebox{0.75}{\usym{2613}}& \scalebox{0.75}{\usym{2613}} & \checkmark & \scalebox{0.75}{\usym{2613}}
    & \scalebox{0.75}{\usym{2613}}& \checkmark  & \scalebox{0.75}{\usym{2613}}
    \\  
    & Wang et al.~\cite{wang2018mashup}
    & \multicolumn{4}{c|}{Knowledge Graph} 
    & 2018
    & \scalebox{0.75}{\usym{2613}} & \scalebox{0.75}{\usym{2613}}& \scalebox{0.75}{\usym{2613}} & \checkmark
    & \scalebox{0.75}{\usym{2613}}& \checkmark  & \scalebox{0.75}{\usym{2613}}
    \\
    & Cao et al.~\cite{cao2024prkg}
    & \multicolumn{4}{c|}{Knowledge Graph + Pre-training}
    & 2024
    & \checkmark & \scalebox{0.75}{\usym{2613}}& \scalebox{0.75}{\usym{2613}} & \checkmark
    & \scalebox{0.75}{\usym{2613}}& \checkmark  & \scalebox{0.75}{\usym{2613}}
    \\
    & Tang et al.~\cite{tang2024light}
    & \multicolumn{4}{c|}{Heterogeneous Hypergraph}
    & 2024
    & \scalebox{0.75}{\usym{2613}} & \checkmark & \scalebox{0.75}{\usym{2613}} & \scalebox{0.75}{\usym{2613}}
    & \scalebox{0.75}{\usym{2613}}& \checkmark  & \scalebox{0.75}{\usym{2613}}
    \\
    \midrule
    \multirow{10}*{\makecell{Dynamic \\ Network \\ Information \\Based}}
    & Zhong et al. ~\cite{zhong2014time}
    & \multicolumn{4}{c|}{LDA} 
    & 2014
    & \checkmark & \scalebox{0.75}{\usym{2613}} & \scalebox{0.75}{\usym{2613}} & \scalebox{0.75}{\usym{2613}}
    & \checkmark& \scalebox{0.75}{\usym{2613}} & \scalebox{0.75}{\usym{2613}}
    \\  
    & Gao et al. ~\cite{gao2017service}
    & \multicolumn{4}{c|}{DC-SeCo-LDA} 
    & 2017
    & \checkmark & \scalebox{0.75}{\usym{2613}} & \scalebox{0.75}{\usym{2613}} & \scalebox{0.75}{\usym{2613}}
    & \checkmark& \scalebox{0.75}{\usym{2613}} & \scalebox{0.75}{\usym{2613}}
    \\  
    & Gao et al. ~\cite{gao2019discovery}
    & \multicolumn{4}{c|}{DC-SeCo-LDA} 
    & 2019
    & \checkmark & \scalebox{0.75}{\usym{2613}} & \scalebox{0.75}{\usym{2613}} & \scalebox{0.75}{\usym{2613}}
    & \checkmark& \scalebox{0.75}{\usym{2613}} & \scalebox{0.75}{\usym{2613}}
    \\  
    & Cao et al. ~\cite{cao2017integrated}
    & \multicolumn{4}{c|}{EDC-SeCo-LDA } 
    & 2017
    & \checkmark & \scalebox{0.75}{\usym{2613}} & \scalebox{0.75}{\usym{2613}} & \scalebox{0.75}{\usym{2613}}
    & \checkmark& \scalebox{0.75}{\usym{2613}} & \scalebox{0.75}{\usym{2613}}
    \\  
    & Huang et al. ~\cite{huang2014recommendation}
    & \multicolumn{4}{c|}{-} 
    & 2014
    & \scalebox{0.75}{\usym{2613}}& \checkmark & \scalebox{0.75}{\usym{2613}} & \scalebox{0.75}{\usym{2613}}
    & \scalebox{0.75}{\usym{2613}} & \scalebox{0.75}{\usym{2613}} & \checkmark
    \\  
    & Adeleye et al. ~\cite{adeleye2020complex}
    & \multicolumn{4}{c|}{Barabási-Albert Complex Network Model} 
    & 2020
    & \scalebox{0.75}{\usym{2613}}& \scalebox{0.75}{\usym{2613}} & \checkmark & \scalebox{0.75}{\usym{2613}}
    & \scalebox{0.75}{\usym{2613}} & \scalebox{0.75}{\usym{2613}} & \checkmark
    \\  
    & Liu et al. ~\cite{liu2021dysr}
    & \multicolumn{4}{c|}{-} 
    & 2021
    & \scalebox{0.75}{\usym{2613}} & \scalebox{0.75}{\usym{2613}}& \scalebox{0.75}{\usym{2613}} & \checkmark
    & \scalebox{0.75}{\usym{2613}} & \scalebox{0.75}{\usym{2613}} & \checkmark
    \\
    & He et al.~\cite{he2025graph}
    & \multicolumn{4}{c|}{GNN + Dynamic Interest Alignment}
    & 2025
    & \scalebox{0.75}{\usym{2613}} & \scalebox{0.75}{\usym{2613}}& \scalebox{0.75}{\usym{2613}} & \checkmark
    & \scalebox{0.75}{\usym{2613}} & \scalebox{0.75}{\usym{2613}} & \checkmark
    \\
    & Zhou et al.~\cite{zhou2024platform}
    & \multicolumn{4}{c|}{Platform Ecosystem Evolution Model}
    & 2024
    & \scalebox{0.75}{\usym{2613}} & \scalebox{0.75}{\usym{2613}}& \checkmark & \scalebox{0.75}{\usym{2613}}
    & \scalebox{0.75}{\usym{2613}} & \scalebox{0.75}{\usym{2613}} & \checkmark
    \\
    & Zhong et al.~\cite{zhong2026popularity}
    & \multicolumn{4}{c|}{Dynamic API Correlation Graph}
    & 2026
    & \scalebox{0.75}{\usym{2613}} & \scalebox{0.75}{\usym{2613}} & \checkmark & \checkmark
    & \scalebox{0.75}{\usym{2613}} & \checkmark & \checkmark
    \\
    \bottomrule
\end{tabular}}
\end{table*}


\subsection{Static network information-based service ecosystem representation}
\label{sec:static_rep}


The representation of a service ecosystem using static network information can be categorized by the static attributes of service nodes and the topological characteristics of the network within the ecosystem. One common static information is the service information, which uses natural language to describe the service functionality in the service ecosystem \cite{zhang2019mining}.

Regarding service description information, Cao et al.~\cite{cao2019qos} used Mashup and Web API description documents, together with their interrelationships, as inputs for topic models. This method captures the relationships between Mashups and services, thus enabling a more precise representation of a service's key features. However, the brief service descriptions pose challenges for effective clustering. Yang et al. \cite{yang2021web} addressed the problem of feature sparsity in short text by expanding word vectors and utilizing topic modeling tailored to brief textual data.

Since modeling a service ecosystem as a complex network, the easiest way to obtain network structure information is through the network topology. Lyu et al.~\cite{lyu2014three} extended the concepts of a service network by introducing the multilayer network and analyzed its structural characteristics. However, they mainly focus on the topological structure information. Liu et al.~\cite{liu2023data} further considered business-level information and proposed the multilayer network-based service ecosystem model (MSEM) to represent the ecosystem, introducing events driving ecosystem evolution. To capture more complex network characteristics, Adeleye et al.~\cite{adeleye2019fitness} used the Bianconi-Barabási complex network\cite{bianconi2001bose,bianconi2001competition} model to construct a service ecosystem, capturing popularity information by the random walk algorithm based on popularity and fitness. 


Recent studies have further advanced static representation by introducing richer graph-based structures. Cao et al.~\cite{cao2024prkg} combine pre-training representations with a service knowledge graph to jointly encode textual semantics, discrete service attributes, and higher-order relations among mashups and services, extending the knowledge graph-based representation paradigm exemplified by Wang et al.~\cite{wang2018mashup}. Tang et al.~\cite{tang2024light} model the mashup-service ecosystem as a heterogeneous hypergraph and apply contrastive learning to capture high-order dependencies among services and mashups, illustrating how static representation is evolving from shallow feature extraction toward structure-aware and relation-aware graph representation learning.

Service descriptions and network topology are fundamental elements through which existing research represents complex networks using static information within service ecosystems. Complex network theories can potentially integrate additional modalities of node attributes, extend beyond textual descriptions, consider heterogeneous or weighted edges, and incorporate dynamic interactions among nodes, communities, and multi-layer networks. This approach could enhance the utilization of static information in representing service ecosystems.

\subsection{Dynamic network information-based service ecosystem representation}
\label{sec:dynamic_rep}

Static network information-based representations of service ecosystems fall short in accurately depicting evolutionary trends due to the omission of temporal information, compared with dynamic information-based network representation methods. 

Zhong et al.~\cite{zhong2014time} found that services often evolve over time, and existing representation methods do not consider the evolution information of service ecosystems. Therefore, they propose a representation method based on LDA~\cite{lda2003} and time series to capture the service ecosystem's historical usage information. Building on this, Gao et al.~\cite{gao2017service} further improved the topic model and proposed the DC-SeCo-LDA model, constructing an evolution graph of the ecosystem to represent the temporal evolution information of the service ecosystem. 

Another type of information that undergoes changes as the service ecosystem evolves is the network topology information. Using a dynamic graph to represent the service ecosystem has become a common method, as it combines the dynamic topology information. Huang et al.~\cite{huang2014recommendation} proposed a network sequence model to define the evolution of the service system, which represents the service ecosystem by quantifying the topological characteristics at different times. Instead of utilizing a simple network model, Adeleye et al.~\cite{adeleye2020complex} employed the advanced Barabási-Albert complex network model to construct an evolving web service network and utilize dynamic changes of the network topology to represent the service ecosystem. Liu et al.~\cite{liu2021dysr} considered external evolution demand information and conducted representation learning based on dynamic graphs. They use a transformation function to bridge the gap between representation and demand, leading to a more accurate representation of the service ecosystem.


More recent work has enriched dynamic representation by modeling not only temporal topology changes but also the continuous alignment between evolving users, services, and demands. He et al.~\cite{he2025graph} propose an incremental service recommendation framework that updates user interests through graph neural network fine-tuning and dynamic interest alignment, reducing catastrophic forgetting while preserving long-term preference information. From a broader ecosystem perspective, Zhou et al.~\cite{zhou2024platform} model platform ecosystem evolution through dynamic supply and matching processes among platforms, users, services, and demands, revealing how market mechanisms drive structural changes in the ecosystem.

Zhong et al.~\cite{zhong2026popularity} use a dynamic simulation and an API correlation graph to study popularity bias in API composition recommendation. Their results show that previous recommendations affect the later exposure and selection of APIs. Dynamic representations should therefore describe not only structural changes, but also the feedback caused by service recommendation and selection.

Dynamic representation methods concentrate on service background information and network topology evolution. From a complex network perspective, this entails tracking time-varying node attributes and the evolution of edges and network structure. Further integration of complex network theory could involve achieving tighter coupling of node attribute co-evolution and topological structures, along with applying advanced analytical techniques such as temporal network analysis and dynamic community detection, to represent the dynamic characteristics of service ecosystems more comprehensively.

The representation methods introduced in this section provide the structural and behavioral foundation for the subsequent three evolutionary stages: they enable the discovery and prediction of evolutionary trends (Section~\ref{sec:Evolution_Perception}), support the formulation of evolutionary strategies such as service composition and migration (Section~\ref{sec_Evolution_Implementation}), and underpin the quantitative evaluation of ecosystem properties (Section~\ref{sec_Evolution_Evaluation}).

%% file: section/sec_perception.tex
\section{Evolution Perception}
\label{sec:Evolution_Perception} 

\begin{figure*}[htbp]
    \centering
    \includegraphics[width=0.9\textwidth]{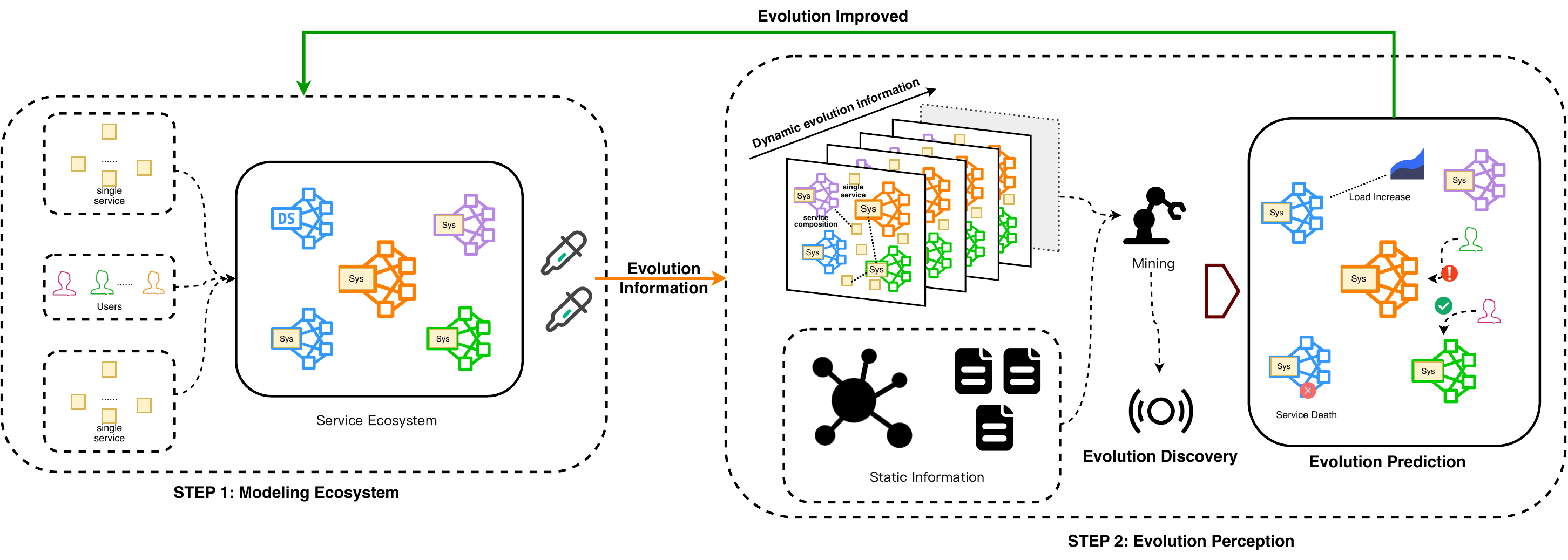}
    \caption{Workflow of service ecosystem evolution perception}
    \label{fig_erceptionworkflow}
\end{figure*}

Evolution discovery and evolution prediction are two stages of evolution perception, as illustrated in Figure~\ref{fig_erceptionworkflow}. Step 1, modeling the ecosystem, involves integrating services and users into a complex network. In Step 2, evolution discovery analyzes past and present data to understand the drivers and patterns of evolution, while evolution prediction focuses on forecasting future trends to guide responses to potential changes. Understanding the causes of ecosystem evolution is essential as it helps identify the reasons behind service ecosystem evolution, anticipate possible shifts, and provide a foundation for evolutionary decisions. 

\subsection{Service ecosystem evolution discovery} 

Evolution discovery involves detecting forthcoming evolution actions in service ecosystems by monitoring historical changes in network structure and dynamic characteristics within the ecosystem. This approach enables the analysis of evolution trends and key driving factors. Based on the type of monitoring information, evolution discovery methods can be categorized into those grounded in network topological structure, network dynamics models, and multi-layer network discovery.

\begin{itemize}
    \item \textbf{Topological structure-based evolution discovery~\cite{maamar2011using, adeleye2019fitness}}: identifies and tracks interdependencies, interaction patterns, and behavior patterns among services through analyzing the topological structure of service networks to improve service discovery accuracy.
    \item \textbf{Network dynamics-based evolution discovery~\cite{6928899, adeleye2020complex}}: reveals how the growth and decay of nodes and edges in complex service ecosystems impact overall behavior, aiding in the exploration of service ecosystem evolution.
    \item \textbf{Multi-layer networks-based evolution discovery~\cite{liu2020novel,zhou2022sle2}}: captures complex relationships and interactions across different levels, efficiently discovering the driving forces and mechanisms of service ecosystem evolution.
\end{itemize}

\subsubsection{Topological structure-based evolution discovery}

In service ecosystems based on complex network representations, network topology facilitated the analysis and tracking of interdependence, interaction patterns, and behavior patterns among services to advance the discovery of service evolution. Acknowledging that considering services as isolated elements could decrease service discovery accuracy~\cite{maamar2011using}, Fallatah et al.~\cite{fallatah2014social} considered the interaction information between users and services participating in the service ecosystem, further expanding the global service network into a social network. They analyzed the topological structure in the social network and discovered possible future interactions between users and services.

Rather than simply expanding the service system by adding users, Chen et al.~\cite{chen2013constructing} proposed a novel framework: a global social service network. This network extended the existing service ecosystem into an open network and utilized specific principles for connecting social services, thereby enhancing both the speed and quality of discovery through a larger topological structure. In contrast to advancing evolution discovery by expanding network topology, Adeleye et al.~\cite{adeleye2019fitness} focused on leveraging existing topological information. They employed a complex network model and assessed the fitness of Web APIs using the random walk algorithm. Subsequently, they developed a continuously evolving social network framework based on the topology's attribute information to improve the completeness of service discovery.


Recent work has further advanced topological evolution discovery through graph-based anomaly localization on complex microservice interaction networks. Tao et al.~\cite{tao2024diagnosing} construct a heterogeneous propagation graph to capture the causal relationships between calls and microservices, and then identify culprit services through a heterogeneity-oriented random walk, achieving 94.1\% top-3 accuracy on real-world cases. Li et al.~\cite{li2025tracedae} further build a Service Trace Graph and combine GAT-based structural encoding with LSTM-based temporal modeling to detect response and invocation anomalies with F1-scores above 0.92. Complementing these runtime monitoring approaches, Cerny et al.~\cite{cerny2025analyzing} develop a static code analysis tool that detects breaking changes and anti-patterns across independently managed microservice codebases, identifying evolution risks before deployment.

Current approaches capture interdependencies and interaction patterns by either expanding the network or directly analyzing topological features, facilitating service discovery and evolution understanding.

\subsubsection{Network dynamics-based evolution discovery}

Network dynamics models examine the growth and decay of nodes and edges to reveal how individual interactions within complex service ecosystems affect overall behavior. These models aid in exploring and discovering service ecosystem evolution.

Aligning with the growth, contraction, merging, splitting, birth, and death layers in the network dynamic model~\cite{palla2007quantifying}, Wang et al.~\cite{6928899} categorized the dynamic processes of service ecosystem evolution into compatibility, transition, split-map, and merge-map, thereby uncovering the underlying causes of service ecosystem evolution. Lu et al.~\cite{lu2020dcem} investigated service evolution and its dynamic behavior based on evolution patterns, particularly focusing on how these evolutionary behaviors could ensure the normal operation of the service ecosystem through mechanisms of automatic self-replacement and self-verification in the event of service failures. To further explore dynamic models of complex networks, Adeleye et al.~\cite{adeleye2020complex} represented the service ecosystem as a Barabási-Albert complex network, dynamically growing the network by publishing both nodes (WebAPIs) and edges (social connections). Additionally, they customized a search method to swiftly respond to complex evolutionary information.

Current work discovers evolution patterns and mechanisms by tracking dynamic events, effectively bridging micro-level element changes with macro-level ecosystem phenomena.

\subsubsection{Multi-layer networks-based evolution discovery} 

Multi-layer network models offer a more comprehensive understanding of the complex relationships and interactions between different levels of the ecosystem. This approach aids in discovering the driving forces and mechanisms of ecosystem evolution, thereby providing a solid foundation for decision-making.

In contrast to the process of identifying individual services via social networks, Brodka et al.~\cite{brodka2013ged} introduced the concept of groups for modeling the service ecosystem within social networks. They employed a method to discover multi-layer network evolution in order to identify seven patterns of individual evolution based on change indicators, a process termed group evolution discovery. Similarly, Liu et al.~\cite{liu2020novel} expanded on this approach by constructing a dynamic evolution trajectory of the service ecosystem using a multi-layer network model. Different from the traditional web service ecosystem, Zhou et al.~\cite{zhou2022sle2} applied the multi-layer network evolution algorithm to the cloud manufacturing service ecosystem and proposed an enhanced model (SLE2) via the same multi-layer network-based method to discover evolution factors. This model also reconstructed the individual, organizational, and social levels to discover the service ecosystem evolution.

Current research utilizes multi-layer networks to identify evolution patterns of individuals or groups and track their dynamic trajectories.

\subsection{Service ecosystem evolution prediction} 

Evolution prediction involves predicting potential changes and evolutionary trends within the service ecosystem. This process typically employs methods such as statistics, machine learning, and artificial intelligence, utilizing historical data and existing knowledge to construct predictive models. 

\begin{itemize}
    \item \textbf{Time series-based evolution prediction~\cite{liu2022community, li2018role}}: analyzes historical data to predict trends and patterns, providing deep insights into the mechanisms driving ecosystem changes over time.
    \item \textbf{Machine learning-based evolution prediction~\cite{wang2016prediction, zhu2022software}}: is increasingly popular in the study of service ecosystem evolution, where algorithms are trained to identify rules in data for making predictions.
    \item \textbf{Pattern recognition-based evolution prediction~\cite{wang2021self,zhou2020measurement}}: aims to predict specific patterns or structures within historical data in service ecosystems to flexibly adjust services.
\end{itemize}

\subsubsection{Time series-based evolution prediction} 

Given the dynamic nature of service ecosystem evolution, time series-based prediction methods are extensively utilized in evolution prediction. These approaches mainly analyze historical data to predict trends and patterns, thereby providing deep insights into the driving mechanisms behind ecosystem changes over time.

In distributed service ecosystems, Chaturvedi et al.~\cite{chaturvedi2020service} employed time series-based evolution prediction methods to identify and extract change information from two evolving versions of the distributed system. This evolution change information was fed into a service evolution analysis model to facilitate further evolution prediction. To address the issue of insufficient predictive accuracy, Liu et al.~\cite{liu2022community} and other researchers adopted a complex representation-based service ecosystem approach to refine prediction granularity at the community level. They developed a time series-based evolution prediction method and trained a model to forecast the evolution of service communities using sequences of community evolution.

Unlike traditional time-series-based prediction methods, Li et al.~\cite{li2018role} quantified the structure of dynamic networks through the use of roles to develop a role model of the dynamic network. They subsequently transformed the evolution prediction into a multi-objective regression problem, utilizing historical time-series data as training data to forecast the potential role distribution.

Current research applies time-series analysis at various granularities (versions, service communities, network roles) to predict future distributions.

\subsubsection{Machine learning-based evolution prediction} 

Machine learning methods have become increasingly popular in various domains, including the study of service ecosystem evolution. These techniques train algorithms to identify patterns and rules in data, which can then be used for making predictions.

Early neural-network models for evolution prediction were relatively simple. Wang et al.~\cite{wang2016prediction} used a multi-layer perceptron (MLP), a type of artificial neural network (ANN), to develop an evolution prediction method. They gathered training data from quality indicators of previous versions of six web services and successfully predicted the evolution of web interfaces, which reduced the workload associated with adapting future application versions. Tan et al.~\cite{tan2021method} proposed a hybrid approach that combined Q-learning with deep learning. They introduced an evolutionary deep-Q-learning algorithm for BPaaS reconstruction to predict business process evolution, enabling finer-grained prediction of quality relevance in business process reengineering. Zhu et al.~\cite{zhu2022software} expanded the application of deep neural networks in service prediction by employing a CNN model to forecast service evolution.

The prediction stage has also begun to benefit from graph-based learning models that explicitly encode evolving service interactions. Huang et al.~\cite{huang2025qos} address QoS prediction for component services in 5G networks through a graph-based deep reinforcement learning framework, where a spatio-temporal recurrent graph attention network is integrated with DDPG to model user mobility and service interaction dynamics. This study demonstrates that recent prediction methods are moving beyond conventional deep learning models toward temporal graph learning architectures that better capture the dynamic structure of service ecosystems.

Machine-learning-based prediction has been applied to forecast interface changes, reconstruct business processes, and manage overall ecosystem changes.


\subsubsection{Pattern recognition-based evolution prediction}
Pattern recognition-based prediction methods aimed to predict specific patterns or structures within the historical data of service ecosystems. To address the frequent occurrence of runtime faults during the evolution of service ecosystems, Wang et al.~\cite{wang2021self} proposed a self-adaptive distributed knowledge-based evolution model. This model integrated historical evolution patterns from service providers into the prediction framework to anticipate potential faults and adaptively adjust services.

Another common evolution pattern within service ecosystems was the refactoring pattern. Stocker et al.~\cite{stocker2021code} investigated the patterns that arose during API evolution in microservices, as conducted by software architects and API developers. Their research offered a catalog of potential patterns that could emerge during API evolution, which served as a foundation for evolution prediction. To enhance the quality of service ecosystems based on predicted evolution patterns, Zhou et al.~\cite{zhou2020measurement} employed two indicators, package cohesion and coupling. These indicators were used to improve the prediction of service ecosystem evolution quality and guide the efficient execution of automatic refactoring.

Current research predicts future behavior by identifying specific patterns within historical evolution data.

%% file: section/sec_implementation.tex
\section{Evolution Implementation}
\label{sec_Evolution_Implementation}
\begin{figure*}[htbp]
    \centering
    \includegraphics[width=0.9\textwidth]{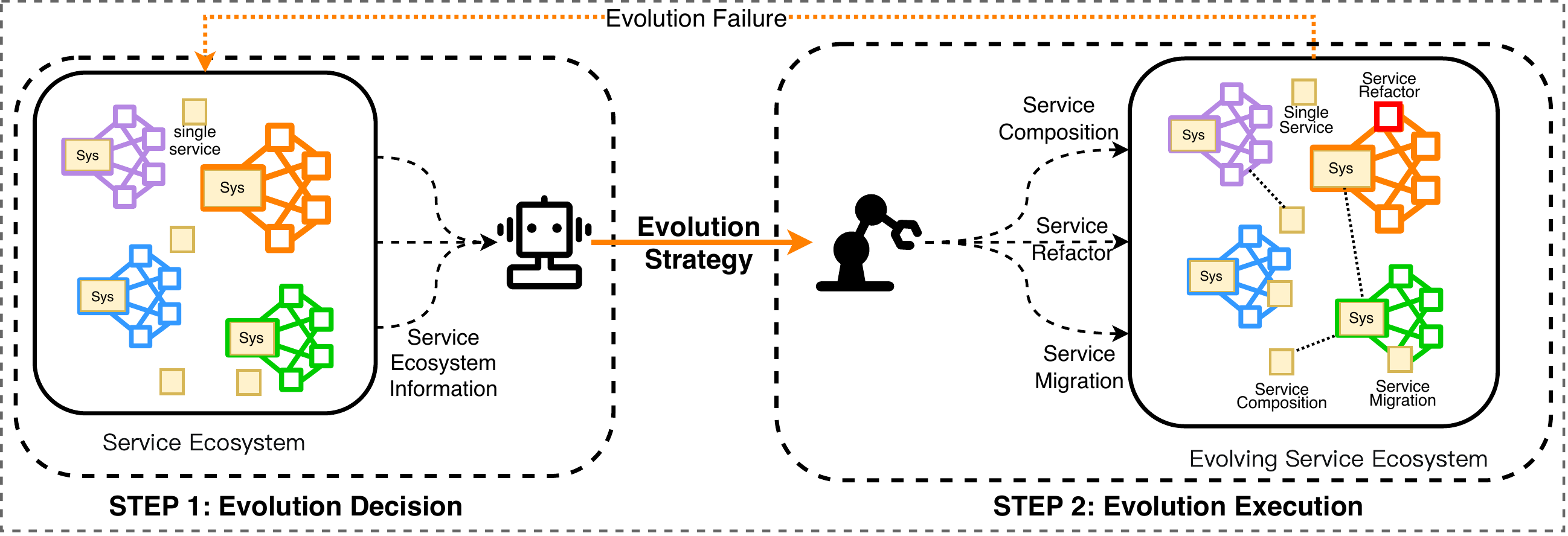}
    \caption{Workflow of service ecosystem evolution implementation illustrating the two-stage process: ``Evolution Decision'' and ``Evolution Execution''.}
    \label{fig_impleworkflow}
\end{figure*}

The evolution implementation stage makes well-informed decisions based on the findings of the evolution perception stage and applies these decisions to the service ecosystem.
This stage can be divided into two substages: evolution decision and evolution execution. Figure~\ref{fig_impleworkflow} shows the overall workflow. Step 1 (Evolution Decision) evaluates the initial service ecosystem information to formulate an evolution strategy, while Step 2 (Evolution Execution) applies this strategy through service refactoring (a structural form of adaptation), service composition, and service migration, thereby transforming the service ecosystem. Table~\ref{table_evolution_implementataion} summarizes representative studies on evolution implementation from a complex-network perspective.

\subsection{Evolution decision}

The increased quantity and diversity of services within service ecosystems pose significant challenges for effective service management and refactoring~\cite{li2011research,briscoe2006digital}. 
The primary goal of the evolution decision stage is to utilize the data collected and analyzed from various aspects of the service ecosystem to formulate models or strategies that determine the optimal implementation plan. When conceptualized as a complex network, service ecosystems can be understood as interconnected structures where individual services function as nodes, service interactions form edges, and related services cluster into communities. Evolution decisions for service ecosystems can be categorized into three groups based on drivers: Demand-Driven, Fault-Driven, and Optimization-Driven.


\subsubsection{Demand-Driven evolution decision} 

Demand-driven evolution decisions aim to respond to changing user and business demands, including the requirements derived from those demands, and to enhance user experience. These decisions often involve reconfiguring the network structure by adding new service nodes, modifying existing node attributes, or restructuring the edges between services to better align with evolving needs.

Several approaches utilized complex network concepts in making demand-driven decisions. Briscoe et al.~\cite{briscoe2006digital} modeled service agents as interacting nodes self-organized through SOA principles, treating applications as emergent patterns and establishing foundational network modeling for evolutionary decision-making.
Building on this, Zhou et al.~\cite{zhou2020evolutionary} framed service composition as edge reconstruction in dependency networks, using QoS metrics (response time, reliability, cost) as weighted attributes to drive co-evolutionary algorithms via a multi-population strategy.
At the community level, Ram{\'i}rez et al.~\cite{Ramrez2017EvolutionaryCO} proposed a three-layer network of requirements, supply nodes, and platform mediators. Their SL2 framework applied genetic algorithms across network scales, mutating node attributes while preserving community cohesion through constraint-aware crossover.

Recent research has further enriched demand-driven evolution decisions by jointly considering multiple quality objectives at runtime. Camilli et al.~\cite{camilli2024integrated} propose a self-adaptive framework that evaluates both QoS and security vulnerabilities through runtime architectural models and probabilistic model checking, enabling adaptation decisions that balance service quality and risk reduction. This study extends the demand-driven decision paradigm by incorporating security considerations into the adaptation decision loop.

\subsubsection{Fault-Driven evolution decision} 

Fault-driven evolution decisions focus on maintaining network integrity by repairing faults, restoring services, and improving robustness and disaster recovery measures~\cite{li2011research}. When viewed as a network problem, this involves detecting anomalous nodes or broken edges, diagnosing the propagation patterns of failures through the network, and implementing targeted interventions to restore network functionality.

Service failures propagate as edge disruptions and node degradations, requiring topology-aware recovery.
Aubonnet et al.~\cite{aubonnet2015management} proposed an adaptive composition framework built on self-controlled components (SCC), modeling services and their sequential/parallel/cyclic relations as first-class network entities to enable dynamic reconfiguration when faults propagate along dependency chains.

Lv et al.~\cite{Lv2018EfficientDE} formalized service dependencies as a DAG stored in an inverted index. Their QoS-aware algorithm traverses the graph to locate alternative paths around failed nodes and continuously updates node/edge states to maintain an optimal composition.

Wang et al.~\cite{Wang2020IntegratingRN} combined graph-based workflow modeling with LSTM~\cite{hochreiter1997long} prediction of node quality and Q-Learning, enabling proactive adjustments before fault points occur.

\subsubsection{Optimization-Driven evolution decision} 

Optimization-driven evolution decisions aim to enhance system quality, adaptability, resource utilization, and reduce costs. Distributed data pre-processing technology addresses performance data structure inconsistency, data incompleteness, and local data identity issues. Invasive or non-invasive software service monitoring tools detect interaction behavior, topology structure, and interface status, enabling dynamic system reconfiguration and service migration plans with collaboration and path strategies.

Systemic optimization requires coordinated adaptation across nodes, edges, and communities. Jatoth et al.~\cite{Jatoth2019OptimalFA} demonstrated node-level optimization through AGEGA's genotype adaptation, using QoS constraints as edge rules to preserve critical dependencies while evolving service populations.
Liu et al.~\cite{liu2022community} shifted focus to community dynamics in multi-layer networks, using temporal community detection and motif analysis to anticipate merger/split events and drive proactive reorganization.
Tan et al.~\cite{tan2021method} accelerated convergence via particle-swarm-optimized Q-Learning over a three-layer DAG, modeling particle movements as potential edge-rewiring operations for cross-community reconfiguration.

In summary, demand changes drive topological reorganization through node/edge modifications, faults disrupt edge connectivity requiring localized repairs, and optimizations balance node improvements with structural preservation.

\subsection{Evolution execution}
The evolution execution stage puts into practice the strategies specified in the evolution decision-making stage to ensure the proper evolution of the service ecosystem. Service adaptation, service migration, and service composition emerge as crucial technologies for implementing evolutionary changes in service ecosystems. 
In this survey, service adaptation is used as an umbrella category, while service refactoring is treated as its structural realization through the reorganization of service boundaries, components, or dependencies. Accordingly, Figure~\ref{fig_impleworkflow} and Table~\ref{table_evolution_implementataion} use service refactoring (SR) for this concrete form of adaptation.

\subsubsection{Service Adaptation-based evolution execution}

Service adaptation refers to the dynamic adjustment process of the structure, behavior, or function of services in response to changes in the internal or external environment, with the aim of maintaining or improving the overall performance, reliability, and user satisfaction of the system.
Service adaptation can be conceptualized as modifications to nodes and edges within the service network, affecting both individual service properties and their interaction patterns.

Service adaptation addresses compatibility issues between services at three layers~\cite{Cao2012,Kuang2009}: message interface, behavior protocol, and functional semantics.

At the message interface layer, Benatallah et al.~\cite{Benatallah2005DevelopingAF} developed an adapter-based method that resolves mismatches in interface names, parameters, types, and quantities, modifying node interface properties to enable compatible connections between previously incompatible services.

At the behavior protocol layer, Yellin et al.~\cite{Yellin1997ProtocolSA} introduced FSM-based protocol modeling to describe communication patterns between components, detecting and resolving protocol mismatches that would otherwise disrupt interactions.

At the functional semantics layer, Cao et al.~\cite{cao2013approach} proposed a semantic service flow network-based adaptation method ensuring behavioral compatibility and correct interaction protocols at network boundaries. Dynamic AOP technologies such as SpringAOP~\cite{SpringAOP} enable runtime modification of node functional behavior without disrupting the established network structure.

Recent work has significantly strengthened service adaptation through graph-based decomposition and self-adaptive refactoring. Zhong et al.~\cite{zhong2024refactoring} propose Micro2Micro, the first approach specifically targeting microservice-to-microservice refactoring in support of evolutionary design, optimizing the interdependence between the refactoring part and the remaining system to achieve up to 45\% architecture quality improvement. Xu et al.~\cite{xu2025ggrme} employ a self-supervised Gated Graph Neural Network (GGNN) for automatic microservice extraction through community detection on system dependency graphs. Xing et al.~\cite{xing2025saber} present SABER, a MAPE-K-based closed-loop framework that detects architectural bad smells and dynamically triggers refactoring strategies including service merging, splitting, and adjustment, achieving 95.53\% precision across ten benchmark systems.

\subsubsection{Service Migration-based evolution execution}
Service Migration comprises selecting suitable migration strategies, determining migration targets, and optimizing resource allocation as services move between different environments or platforms within the ecosystem. From a network perspective, migration can be viewed as the relocation of service nodes while preserving their functional roles and connections within the overall network structure.

Taleb et al.~\cite{Taleb2013AnAM} proposed FollowMe Cloud (FMC), which migrates services to track user mobility across network locations, using an analytical model over user mobility patterns, migration strategies, and data-center distributions to optimize node placement and preserve service quality.

Machen et al.~\cite{Machen2016MigratingRA} decomposed migration into infrastructure, application, and instance layers, transferring components independently to reduce downtime.

Papageorgiou et al.~\cite{Papageorgiou2014DecisionSF} proposed a quality-driven decision algorithm that treats non-functional QoS/QoE attributes as weighted node/edge properties, guiding migration to preserve path and community performance.

For mobile environments, Wang et al.~\cite{Wang2011ExploitingMP} incorporated provider mobility into migration via two composition selection methods matching distinct mobility patterns, while Deng et al.~\cite{Deng2016TowardRR} quantified mobility-induced reliability risk on edges to prefer stably connectable services.

\subsubsection{Service Composition-based evolution execution}

Service Composition is the process of combining, coordinating, and integrating multiple independent services into a new, higher-level service that satisfies specific user requirements or business goals. In network terms, this involves creating new functional pathways or subgraphs within the service ecosystem by establishing strategic connections between existing service nodes, effectively generating composite services that leverage the collective capabilities of their components.

For resource-constrained environments, Chen et al.~\cite{Chen2014ADS} proposed a distributed composition algorithm that accounts for heterogeneous computational capacity of service endpoints, distributing the selection process across the network for edge computing and IoT scenarios.

Tan et al.~\cite{tan2021method} modeled services as a three-layer DAG and combined Q-Learning with particle swarm optimization for adaptive composition under changing network conditions.

For community-aware composition, Ram{\'i}rez et al.~\cite{Ramrez2017EvolutionaryCO} employed genetic algorithms and neural networks over a multi-layer model of providers, consumers, and platforms, leveraging social, supply-chain, and dependency relationships at the community level.

In dynamic environments, Wang et al.~\cite{Wang2020IntegratingRN} combined LSTM-based quality prediction with reinforcement learning to anticipate path-level performance and proactively adjust compositions.

Liu et al.~\cite{liujiwei2016} proposed a meta-variable-based technology supporting addition, replacement, and deletion of variability elements, providing controlled topology modifications through dynamic architecture adjustment.

Yang et al.~\cite{yang2025service} propose a dual-layer coupled network model combined with improved particle swarm optimization for cloud manufacturing service composition, where topology analysis through node centrality metrics guides initial service screening. This network-centric composition approach explicitly leverages complex network properties for evolution execution.

Zhang et al.~\cite{zhang2025association} formulate a three-objective service-composition model covering service quality, cost, and service association. An improved NSGA-II algorithm achieves better convergence and solution diversity than the compared methods. However, the method is evaluated only in cloud-manufacturing scenarios with predefined objectives, so its general applicability remains unclear.

In summary, these approaches for evolution execution---service adaptation, migration, and composition---modify the ecosystem by manipulating nodes (adjusting properties, relocating, adding/deleting) and edges (modifying connections, ensuring compatibility, optimizing paths) to achieve specific evolutionary goals.

To comparatively summarize the representative studies across the evolution implementation stage, Table~\ref{table_evolution_implementataion} lists the selected works according to their network classification, methodology, and evolution goal. In the table, EA represents evolutionary algorithm, NN represents neural network, Rule represents rule-based approaches, SC represents service composition, SR represents service refactoring as a structural form of adaptation, and SM represents service migration.

\begin{table*}[!htbp]
\centering
\caption{Summary of the work on evolution implementation}
\label{table_evolution_implementataion}
\footnotesize
\renewcommand{\arraystretch}{1.15}
\setlength{\tabcolsep}{0.8mm}{
\begin{tabular}{c|c|c|ccc|cccc|cccc}
    \toprule
    \multicolumn{2}{c|}{ \multirow{2}*{Related Work} }
    & \multirow{2}*{Year}
    & \multicolumn{3}{c|}{Network Classification}
    & \multicolumn{4}{c|}{Methodology}
    & \multicolumn{4}{c}{Evolution Goal} \\
    \multicolumn{2}{c|}{}&
    & Node & Chain & Community
    & EA & NN & Rule & Others
    & QoS & SC & SR & SM
    \\ \midrule

    \multirow{11}*{\makecell{Evolution \\ Decision}}
    & Briscoe et al.~\cite{briscoe2006digital} & 2006
    & \checkmark & \scalebox{0.75}{\usym{2613}} & \scalebox{0.75}{\usym{2613}}
    & \scalebox{0.75}{\usym{2613}} & \scalebox{0.75}{\usym{2613}} & \scalebox{0.75}{\usym{2613}} & \checkmark
    & \checkmark & \scalebox{0.75}{\usym{2613}}  & \scalebox{0.75}{\usym{2613}}  & \scalebox{0.75}{\usym{2613}}
    \\
    & Zhou et al.~\cite{zhou2020evolutionary} & 2020
    & \checkmark & \scalebox{0.75}{\usym{2613}} & \scalebox{0.75}{\usym{2613}}
    & \checkmark& \scalebox{0.75}{\usym{2613}} & \scalebox{0.75}{\usym{2613}} & \scalebox{0.75}{\usym{2613}}
    & \scalebox{0.75}{\usym{2613}} & \checkmark & \scalebox{0.75}{\usym{2613}}  & \scalebox{0.75}{\usym{2613}}
    \\
    & Jatoth et al.~\cite{Jatoth2019OptimalFA}&2019
    & \checkmark & \scalebox{0.75}{\usym{2613}} & \scalebox{0.75}{\usym{2613}}
    & \checkmark & \scalebox{0.75}{\usym{2613}} & \scalebox{0.75}{\usym{2613}} & \scalebox{0.75}{\usym{2613}}
    & \scalebox{0.75}{\usym{2613}} & \checkmark& \scalebox{0.75}{\usym{2613}}  & \scalebox{0.75}{\usym{2613}}
    \\
    & Aubonnet et al.~\cite{aubonnet2015management} &2015
    & \scalebox{0.75}{\usym{2613}} & \checkmark & \scalebox{0.75}{\usym{2613}}
    & \scalebox{0.75}{\usym{2613}} & \scalebox{0.75}{\usym{2613}} & \checkmark & \scalebox{0.75}{\usym{2613}}
    & \checkmark & \scalebox{0.75}{\usym{2613}} & \scalebox{0.75}{\usym{2613}}  & \scalebox{0.75}{\usym{2613}}
    \\
    & Lv et al.~\cite{Lv2018EfficientDE} &2018
    & \scalebox{0.75}{\usym{2613}} & \checkmark & \scalebox{0.75}{\usym{2613}}
    & \scalebox{0.75}{\usym{2613}} & \checkmark & \scalebox{0.75}{\usym{2613}} & \scalebox{0.75}{\usym{2613}}
    & \checkmark& \scalebox{0.75}{\usym{2613}}& \scalebox{0.75}{\usym{2613}}  & \scalebox{0.75}{\usym{2613}}
    \\
    & Ram{\'\i}rez et al.~\cite{Ramrez2017EvolutionaryCO} &2017
    & \scalebox{0.75}{\usym{2613}}& \scalebox{0.75}{\usym{2613}} & \checkmark
    & \checkmark & \scalebox{0.75}{\usym{2613}} & \scalebox{0.75}{\usym{2613}} & \scalebox{0.75}{\usym{2613}}
    & \scalebox{0.75}{\usym{2613}}& \checkmark & \scalebox{0.75}{\usym{2613}}  & \scalebox{0.75}{\usym{2613}}
    \\
    & Liu et al.~\cite{liu2022community} &2022
    & \scalebox{0.75}{\usym{2613}}& \scalebox{0.75}{\usym{2613}} & \checkmark
    & \checkmark & \scalebox{0.75}{\usym{2613}} & \scalebox{0.75}{\usym{2613}} & \scalebox{0.75}{\usym{2613}}
    & \scalebox{0.75}{\usym{2613}} & \checkmark& \scalebox{0.75}{\usym{2613}}  & \scalebox{0.75}{\usym{2613}}
    \\
    & Zhao et al.~\cite{chenzhaojie2021} & 2021
    & \scalebox{0.75}{\usym{2613}} & \scalebox{0.75}{\usym{2613}}& \scalebox{0.75}{\usym{2613}}
    & \scalebox{0.75}{\usym{2613}}& \checkmark  & \scalebox{0.75}{\usym{2613}} & \scalebox{0.75}{\usym{2613}}
    & \scalebox{0.75}{\usym{2613}} & \checkmark& \scalebox{0.75}{\usym{2613}}  & \scalebox{0.75}{\usym{2613}}
    \\
    & Wang et al.~\cite{Wang2020IntegratingRN} & 2020
    & \scalebox{0.75}{\usym{2613}} & \scalebox{0.75}{\usym{2613}} & \scalebox{0.75}{\usym{2613}}
    & \scalebox{0.75}{\usym{2613}} & \scalebox{0.75}{\usym{2613}} & \scalebox{0.75}{\usym{2613}} & \checkmark
    & \scalebox{0.75}{\usym{2613}} & \checkmark & \scalebox{0.75}{\usym{2613}}  & \scalebox{0.75}{\usym{2613}}
    \\
    & Tan et al.~\cite{tan2021method} & 2021
    & \scalebox{0.75}{\usym{2613}} & \scalebox{0.75}{\usym{2613}}& \scalebox{0.75}{\usym{2613}}
    & \scalebox{0.75}{\usym{2613}} & \checkmark & \scalebox{0.75}{\usym{2613}}& \scalebox{0.75}{\usym{2613}}
    & \scalebox{0.75}{\usym{2613}} & \checkmark & \scalebox{0.75}{\usym{2613}}  & \scalebox{0.75}{\usym{2613}}
    \\
    & Camilli et al.~\cite{camilli2024integrated} & 2024
    & \checkmark & \scalebox{0.75}{\usym{2613}} & \scalebox{0.75}{\usym{2613}}
    & \scalebox{0.75}{\usym{2613}} & \scalebox{0.75}{\usym{2613}} & \checkmark & \scalebox{0.75}{\usym{2613}}
    & \checkmark & \scalebox{0.75}{\usym{2613}} & \scalebox{0.75}{\usym{2613}} & \scalebox{0.75}{\usym{2613}}
    \\
    \midrule

    \multirow{17}*{\makecell{Evolution \\ Execution}}
    & Machen et al.~\cite{Machen2016MigratingRA} & 2016
    & \checkmark& \scalebox{0.75}{\usym{2613}}& \scalebox{0.75}{\usym{2613}}
    & \scalebox{0.75}{\usym{2613}} & \scalebox{0.75}{\usym{2613}} & \scalebox{0.75}{\usym{2613}} & \checkmark
    & \scalebox{0.75}{\usym{2613}} & \scalebox{0.75}{\usym{2613}} & \scalebox{0.75}{\usym{2613}} & \checkmark
    \\
    & Papageorgiou et al.~\cite{Papageorgiou2014DecisionSF} & 2014
    & \checkmark& \scalebox{0.75}{\usym{2613}}& \scalebox{0.75}{\usym{2613}}
    & \scalebox{0.75}{\usym{2613}} & \checkmark & \scalebox{0.75}{\usym{2613}} & \scalebox{0.75}{\usym{2613}}
    & \scalebox{0.75}{\usym{2613}} & \scalebox{0.75}{\usym{2613}} & \checkmark & \scalebox{0.75}{\usym{2613}}
    \\
    & Benatallah et al.~\cite{Benatallah2005DevelopingAF} & 2005
    & \scalebox{0.75}{\usym{2613}} & \checkmark & \scalebox{0.75}{\usym{2613}}
    & \scalebox{0.75}{\usym{2613}}& \scalebox{0.75}{\usym{2613}} & \scalebox{0.75}{\usym{2613}}  & \checkmark
    & \scalebox{0.75}{\usym{2613}}& \scalebox{0.75}{\usym{2613}} & \checkmark & \scalebox{0.75}{\usym{2613}}
    \\
    & Yellin et al.~\cite{Yellin1997ProtocolSA} & 1997
    & \scalebox{0.75}{\usym{2613}} & \checkmark & \scalebox{0.75}{\usym{2613}}
    & \scalebox{0.75}{\usym{2613}} & \scalebox{0.75}{\usym{2613}} & \scalebox{0.75}{\usym{2613}} & \checkmark
    & \scalebox{0.75}{\usym{2613}}& \scalebox{0.75}{\usym{2613}} & \checkmark & \scalebox{0.75}{\usym{2613}}
    \\
    & Cao et al.~\cite{cao2013approach} & 2013
    & \scalebox{0.75}{\usym{2613}} & \checkmark & \scalebox{0.75}{\usym{2613}}
    & \scalebox{0.75}{\usym{2613}} & \scalebox{0.75}{\usym{2613}} & \scalebox{0.75}{\usym{2613}} & \checkmark
    & \scalebox{0.75}{\usym{2613}}& \scalebox{0.75}{\usym{2613}} & \checkmark & \scalebox{0.75}{\usym{2613}}
    \\
    & Ha et al.~\cite{Ha2015AdaptiveVH} & 2015
    & \scalebox{0.75}{\usym{2613}}& \scalebox{0.75}{\usym{2613}} & \checkmark
    & \scalebox{0.75}{\usym{2613}} & \scalebox{0.75}{\usym{2613}} & \scalebox{0.75}{\usym{2613}} & \checkmark
    & \scalebox{0.75}{\usym{2613}}  & \checkmark & \scalebox{0.75}{\usym{2613}} & \scalebox{0.75}{\usym{2613}}
    \\
    & Wang et al.~\cite{Wang2011ExploitingMP} & 2011
    & \scalebox{0.75}{\usym{2613}}& \scalebox{0.75}{\usym{2613}} & \checkmark
    & \scalebox{0.75}{\usym{2613}} & \scalebox{0.75}{\usym{2613}} & \scalebox{0.75}{\usym{2613}} & \checkmark
    & \scalebox{0.75}{\usym{2613}}  & \checkmark & \scalebox{0.75}{\usym{2613}} & \scalebox{0.75}{\usym{2613}}
    \\
    & Deng et al.~\cite{Deng2016TowardRR} & 2016
    & \scalebox{0.75}{\usym{2613}}& \scalebox{0.75}{\usym{2613}} & \checkmark
    & \scalebox{0.75}{\usym{2613}} & \scalebox{0.75}{\usym{2613}} & \scalebox{0.75}{\usym{2613}} & \checkmark
    & \scalebox{0.75}{\usym{2613}}  & \checkmark & \scalebox{0.75}{\usym{2613}} & \scalebox{0.75}{\usym{2613}}
    \\
    & Chen et al.~\cite{Chen2014ADS} & 2014
    & \scalebox{0.75}{\usym{2613}}& \scalebox{0.75}{\usym{2613}} & \checkmark
    & \scalebox{0.75}{\usym{2613}} & \scalebox{0.75}{\usym{2613}} & \scalebox{0.75}{\usym{2613}} & \checkmark
    & \scalebox{0.75}{\usym{2613}}  & \checkmark & \scalebox{0.75}{\usym{2613}} & \scalebox{0.75}{\usym{2613}}
    \\
    & Liu et al.~\cite{liujiwei2016} & 2016
    & \scalebox{0.75}{\usym{2613}}& \scalebox{0.75}{\usym{2613}} & \scalebox{0.75}{\usym{2613}}
    & \scalebox{0.75}{\usym{2613}} & \scalebox{0.75}{\usym{2613}} & \scalebox{0.75}{\usym{2613}} & \checkmark
    & \scalebox{0.75}{\usym{2613}} & \scalebox{0.75}{\usym{2613}}  & \checkmark & \checkmark
    \\
    & Spring AOP~\cite{SpringAOP} & 2002
    & \scalebox{0.75}{\usym{2613}}& \scalebox{0.75}{\usym{2613}} & \scalebox{0.75}{\usym{2613}}
    & \scalebox{0.75}{\usym{2613}} & \scalebox{0.75}{\usym{2613}} & \scalebox{0.75}{\usym{2613}} & \checkmark
    & \scalebox{0.75}{\usym{2613}} & \scalebox{0.75}{\usym{2613}}  & \checkmark & \checkmark
    \\
    & Lu et al.~\cite{lu2020dcem} & 2020
    & \scalebox{0.75}{\usym{2613}}& \scalebox{0.75}{\usym{2613}} & \scalebox{0.75}{\usym{2613}}
    & \scalebox{0.75}{\usym{2613}} & \scalebox{0.75}{\usym{2613}} & \scalebox{0.75}{\usym{2613}} & \checkmark
    & \scalebox{0.75}{\usym{2613}} & \scalebox{0.75}{\usym{2613}}  & \checkmark & \checkmark
    \\
    & Zhong et al.~\cite{zhong2024refactoring} & 2024
    & \scalebox{0.75}{\usym{2613}} & \scalebox{0.75}{\usym{2613}} & \checkmark
    & \checkmark & \scalebox{0.75}{\usym{2613}} & \scalebox{0.75}{\usym{2613}} & \scalebox{0.75}{\usym{2613}}
    & \scalebox{0.75}{\usym{2613}} & \scalebox{0.75}{\usym{2613}} & \checkmark & \scalebox{0.75}{\usym{2613}}
    \\
    & Xu et al.~\cite{xu2025ggrme} & 2025
    & \scalebox{0.75}{\usym{2613}} & \scalebox{0.75}{\usym{2613}} & \checkmark
    & \scalebox{0.75}{\usym{2613}} & \checkmark & \scalebox{0.75}{\usym{2613}} & \scalebox{0.75}{\usym{2613}}
    & \scalebox{0.75}{\usym{2613}} & \scalebox{0.75}{\usym{2613}} & \checkmark & \scalebox{0.75}{\usym{2613}}
    \\
    & Xing et al.~\cite{xing2025saber} & 2025
    & \scalebox{0.75}{\usym{2613}} & \scalebox{0.75}{\usym{2613}} & \checkmark
    & \scalebox{0.75}{\usym{2613}} & \scalebox{0.75}{\usym{2613}} & \checkmark & \scalebox{0.75}{\usym{2613}}
    & \scalebox{0.75}{\usym{2613}} & \scalebox{0.75}{\usym{2613}} & \checkmark & \scalebox{0.75}{\usym{2613}}
    \\
    & Yang et al.~\cite{yang2025service} & 2025
    & \checkmark & \scalebox{0.75}{\usym{2613}} & \scalebox{0.75}{\usym{2613}}
    & \checkmark & \scalebox{0.75}{\usym{2613}} & \scalebox{0.75}{\usym{2613}} & \scalebox{0.75}{\usym{2613}}
    & \scalebox{0.75}{\usym{2613}} & \checkmark & \scalebox{0.75}{\usym{2613}} & \scalebox{0.75}{\usym{2613}}
    \\
    & Zhang et al.~\cite{zhang2025association} & 2025
    & \scalebox{0.75}{\usym{2613}} & \checkmark & \scalebox{0.75}{\usym{2613}}
    & \checkmark & \scalebox{0.75}{\usym{2613}} & \scalebox{0.75}{\usym{2613}} & \scalebox{0.75}{\usym{2613}}
    & \checkmark & \checkmark & \scalebox{0.75}{\usym{2613}} & \scalebox{0.75}{\usym{2613}}
    \\
    \bottomrule
\end{tabular}}
\end{table*}

%% file: section/sec_evaluation.tex
\begin{figure*}[htbp]
    \centering
    \includegraphics[width=0.9\textwidth]{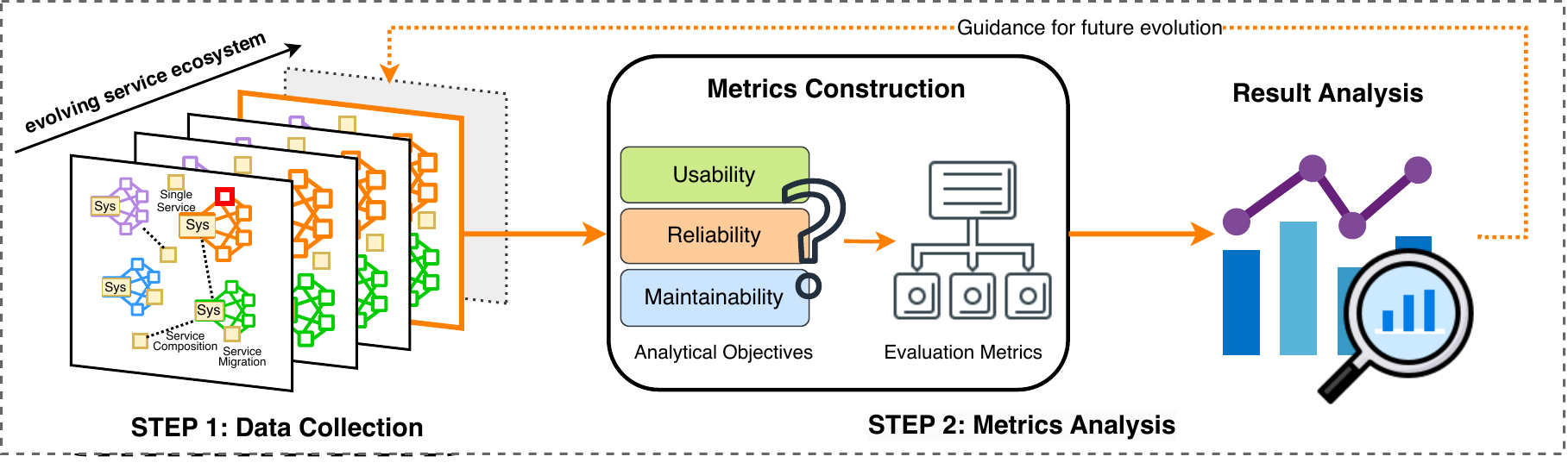}
    \caption{Workflow of evolution evaluation for a service ecosystem.}
    \label{fig_evalworkflow}
\end{figure*}

\section{Evolution Evaluation}
\label{sec_Evolution_Evaluation}

Service ecosystem evolution aims to maintain and enhance service quality. This section reviews evaluation approaches through two steps, as shown in Figure~\ref{fig_evalworkflow}: data collection from appropriate sources and metrics analysis targeting reliability, usability, and maintainability.

\subsection{Data Collection}
We distinguish synthetic datasets, real-world data, and case studies, together with monitoring-based and human-based collection strategies.
\subsubsection{Data Sources}
\textbf{Synthetic Datasets.} Controlled data help isolate specific mechanisms and parameters:
\begin{itemize}
    \item \textit{Randomly Generated Service Models:} \cite{huang2015failure} used feature and service generators with parameterized distributions to evaluate adaptive compositions under functional failures and QoS violations while isolating individual parameters' effects on performance.
    
    \item \textit{Mutant Generation:} \cite{bashari2018self} created fault-seeded program variants to evaluate fault localization and self-healing through test suites with known fault locations.
    
    \item \textit{Synthetic Service Repositories:} \cite{margaris2016improving} tested QoS optimization and parallelization across repositories with varying scales, complexity levels, and composition structures.
\end{itemize}

Synthetic datasets support controlled comparisons of correctness, efficiency, and scalability, although they cannot fully reproduce real-world ecosystem complexity.

\textbf{Real-World Data.} Operational datasets provide more realistic ecosystem behavior:
\begin{itemize}
    \item \textit{Industry Cluster Data:} \cite{jia2020research} analyzed longitudinal automotive-industry data to study competitive and cooperative dynamics among service groups and validate predictive evolution models.
    
    \item \textit{Web API Directory Data:} \cite{adeleye2018constructing} used ProgrammableWeb data to examine service-network evolution, network formation, and emergent properties over time.
    
    \item \textit{Corporate Financial Data:} \cite{xue2020integrative,xue2020value} used operational records from major companies to calibrate parameters and validate theoretical models against market performance.
\end{itemize}

Despite noise and confounding factors, real-world data improve external validity and reveal long-term patterns under realistic conditions.

\textbf{Case Studies.} Scenario-based evaluations balance experimental control with practical context:
\begin{itemize}
    \item \textit{Self-constructed Test Scenarios:} \cite{huang2015failure} simulated specific failures to evaluate recovery mechanisms and compare responses to different fault types.
    
    \item \textit{Classic Application Scenarios:} \cite{alferez2017achieving,subramanian2008enhancement} used online shopping and travel services to validate adaptation frameworks in recognizable domain settings.
\end{itemize}

Case studies therefore connect controlled experiments with practical, domain-specific evaluation.

Together, these sources trade experimental control for external validity: synthetic data isolate mechanisms, real-world data capture ecosystem complexity, and case studies connect both perspectives.

\subsubsection{Data Collection Strategies}

\textbf{Monitoring-Based Collection.} These approaches capture runtime behavior:
\begin{itemize}
    \item \textit{Sensor Monitoring:}~\cite{alferez2017achieving} tracked availability and execution time using ping/echo methods to evaluate runtime adaptation and temporal variations in service quality.
    
    \item \textit{Event Log Collection:} \cite{asim2018security} gathered detailed load-test event logs to evaluate compliance monitoring and anomaly detection under stress.
\end{itemize}

\textbf{Human-Based Collection.} These approaches capture practitioner and user assessments:
\begin{itemize} 
    \item \textit{Crowdsourcing Platforms:} \cite{alkalbani_quality_2018} collected SaaS QoS ratings, comments, and questionnaire responses from minimally trained participants.
    \item \textit{Interview-Based Research:} \cite{taibi_definition_2018,jaafar_analyzing_2017} surveyed microservice experts about maintenance challenges, practices, and lessons from development experience.
\end{itemize}

\subsection{Metrics Analysis}
This subsection presents a framework for quantifying service ecosystem evolution through reliability, usability, and maintainability metrics. We identify key indicators, discuss their significance in measuring evolutionary success, and examine analytical methods for interpreting these measurements to guide ecosystem development.
To comparatively summarize the representative studies across the three evaluation dimensions, Table~\ref{table:evolution_evaluation} lists the selected works according to their methodology, modeling object, and network classification. In the table, SC represents service configuration, SS represents service substitution, FD represents fault discovery, MA represents model analysis, SSG represents specific service composition, and GSE represents general service ecosystem.

\begin{table*}[!t]
  \centering
  \caption{Summaries and comparisons of evolution evaluation}
  \label{table:evolution_evaluation}
  \footnotesize
  \renewcommand{\arraystretch}{1.15}
  \setlength{\tabcolsep}{0.8mm}{
  \begin{tabular}{cc|c|cccc|cc|cc}
  \toprule
  \multicolumn{2}{c|}{\multirow{2}{*}{Related Work}} &
    \multirow{2}{*}{Year} &
    \multicolumn{4}{c|}{Methodology} &
    \multicolumn{2}{c|}{Modeling Object} &
    \multicolumn{2}{c}{Network classification} \\ \cline{4-11}
  \multicolumn{2}{c|}{} &
     &
     SC & SS & FD & MA &
     SSG&GSE&
     Node&Community\\ \midrule
  \multicolumn{1}{c|}{\multirow{13}{*}{Reliability}} &
      Huang et al.~\cite{huang2015failure}&2015
      & \checkmark&\scalebox{0.75}{\usym{2613}}&\scalebox{0.75}{\usym{2613}}&\scalebox{0.75}{\usym{2613}}
      & \checkmark& \scalebox{0.75}{\usym{2613}}
      & \checkmark& \scalebox{0.75}{\usym{2613}}\\

      \multicolumn{1}{c|}{}& Bashari et al.~\cite{bashari2018self}&2018
      & \checkmark&\scalebox{0.75}{\usym{2613}}&\scalebox{0.75}{\usym{2613}}&\scalebox{0.75}{\usym{2613}}
      & \checkmark& \scalebox{0.75}{\usym{2613}}
      & \checkmark & \scalebox{0.75}{\usym{2613}}\\

      \multicolumn{1}{c|}{}& Alferez et al.~\cite{alferez2017achieving}&2017
      & \checkmark&\scalebox{0.75}{\usym{2613}}&\scalebox{0.75}{\usym{2613}}&\scalebox{0.75}{\usym{2613}}
      & \checkmark& \scalebox{0.75}{\usym{2613}}
      & \checkmark& \scalebox{0.75}{\usym{2613}}\\

      \multicolumn{1}{c|}{}& Margaris et al.~\cite{margaris2016improving}&2016
      &\scalebox{0.75}{\usym{2613}}& \checkmark&\scalebox{0.75}{\usym{2613}}&\scalebox{0.75}{\usym{2613}}
      & \checkmark& \scalebox{0.75}{\usym{2613}}
      & \checkmark& \scalebox{0.75}{\usym{2613}}\\

      \multicolumn{1}{c|}{}& Lu et al.~\cite{lu2020dcem}&2020
      & \scalebox{0.75}{\usym{2613}}& \checkmark& \scalebox{0.75}{\usym{2613}}& \scalebox{0.75}{\usym{2613}}
      & \checkmark& \scalebox{0.75}{\usym{2613}}
      &{\scalebox{0.75}{\usym{2613}}}& \checkmark\\

      \multicolumn{1}{c|}{}& Sun et al.~\cite{sun2018fault}&2018
      & \scalebox{0.75}{\usym{2613}}& \scalebox{0.75}{\usym{2613}}& \checkmark& \scalebox{0.75}{\usym{2613}}
      & \checkmark& \scalebox{0.75}{\usym{2613}}
      & \checkmark& \scalebox{0.75}{\usym{2613}}\\

      \multicolumn{1}{c|}{}& Asim et al.~\cite{asim2018security}&2018
      & \scalebox{0.75}{\usym{2613}}& \scalebox{0.75}{\usym{2613}}& \checkmark& \scalebox{0.75}{\usym{2613}}
      & \checkmark& \scalebox{0.75}{\usym{2613}}
      & \checkmark& \scalebox{0.75}{\usym{2613}}\\

      \multicolumn{1}{c|}{}& Jia et al.~\cite{jia2020research}&2020
      & \scalebox{0.75}{\usym{2613}}& \scalebox{0.75}{\usym{2613}}& \scalebox{0.75}{\usym{2613}}& \checkmark
      & \scalebox{0.75}{\usym{2613}}& \checkmark
      &{\scalebox{0.75}{\usym{2613}}}& \checkmark\\

      \multicolumn{1}{c|}{}& Xue et al.~\cite{xue2020integrative}&2020
      & \scalebox{0.75}{\usym{2613}}& \scalebox{0.75}{\usym{2613}}& \scalebox{0.75}{\usym{2613}}& \checkmark
      & \scalebox{0.75}{\usym{2613}}& \checkmark
      &{\scalebox{0.75}{\usym{2613}}}& \checkmark\\

      \multicolumn{1}{c|}{}& Xue et al.~\cite{xue2020value}&2020
      & \scalebox{0.75}{\usym{2613}}& \scalebox{0.75}{\usym{2613}}& \scalebox{0.75}{\usym{2613}}& \checkmark
      & \scalebox{0.75}{\usym{2613}}& \checkmark
      &{\scalebox{0.75}{\usym{2613}}}& \checkmark\\

      \multicolumn{1}{c|}{}& Adeleye et al.~\cite{adeleye2018constructing}&2018
      & \scalebox{0.75}{\usym{2613}}& \scalebox{0.75}{\usym{2613}}& \scalebox{0.75}{\usym{2613}}& \checkmark
      & \scalebox{0.75}{\usym{2613}}& \checkmark
      &{\scalebox{0.75}{\usym{2613}}}& \checkmark\\

      \multicolumn{1}{c|}{}& Pham et al.~\cite{pham2024baro}&2024
      & \scalebox{0.75}{\usym{2613}}& \scalebox{0.75}{\usym{2613}}& \checkmark& \scalebox{0.75}{\usym{2613}}
      & \checkmark& \scalebox{0.75}{\usym{2613}}
      & \checkmark& \scalebox{0.75}{\usym{2613}}\\

      \multicolumn{1}{c|}{}& Zhang et al.~\cite{zhang2025network}&2025
      & \scalebox{0.75}{\usym{2613}}& \scalebox{0.75}{\usym{2613}}& \scalebox{0.75}{\usym{2613}}& \checkmark
      & \scalebox{0.75}{\usym{2613}}& \checkmark
      & \checkmark& \checkmark\\ \midrule
  \multicolumn{1}{c|}{\multirow{10}{*}{Usability}} &
    Goeb and Lochmann \cite{goeb_software_2011} &
    2011 &
    \scalebox{0.75}{\usym{2613}} & \scalebox{0.75}{\usym{2613}} & \scalebox{0.75}{\usym{2613}} & \checkmark & \scalebox{0.75}{\usym{2613}} & \checkmark &
    {\checkmark} & \usym{2613} \\
  \multicolumn{1}{c|}{} &
    Phipathananunth and Bunyakiati \cite{phipathananunth_synthetic_2018} &
    2018 &
    \scalebox{0.75}{\usym{2613}} & \scalebox{0.75}{\usym{2613}} & \scalebox{0.75}{\usym{2613}} & \checkmark & \scalebox{0.75}{\usym{2613}} & \checkmark &
    {\checkmark} & \usym{2613} \\
  \multicolumn{1}{c|}{} &
    Alkalbani and Hussain \cite{alkalbani_quality_2018} &
    2018 &
    \scalebox{0.75}{\usym{2613}} & \scalebox{0.75}{\usym{2613}} & \scalebox{0.75}{\usym{2613}} & \scalebox{0.75}{\usym{2613}} & \scalebox{0.75}{\usym{2613}} & \checkmark &
    {\checkmark} & \usym{2613} \\
  \multicolumn{1}{c|}{} &
    Stévant et al.~\cite{stevant_optimizing_2018} &
    2018 &
    \scalebox{0.75}{\usym{2613}} & \scalebox{0.75}{\usym{2613}} & \scalebox{0.75}{\usym{2613}} & \checkmark & \scalebox{0.75}{\usym{2613}} & \checkmark &
    {\checkmark} & \usym{2613} \\
  \multicolumn{1}{c|}{} &
    Gribaudo et al.~\cite{gribaudo_performance_2017} &
    2017 &
    \scalebox{0.75}{\usym{2613}} & \scalebox{0.75}{\usym{2613}} & \scalebox{0.75}{\usym{2613}} & \checkmark & \scalebox{0.75}{\usym{2613}} & \checkmark &
    {\usym{2613}} & \checkmark \\
  \multicolumn{1}{c|}{} &
    Avritzer et al.~\cite{avritzer_quantitative_2018} &
    2018 &
    \scalebox{0.75}{\usym{2613}} & \scalebox{0.75}{\usym{2613}} & \scalebox{0.75}{\usym{2613}} & \checkmark & \scalebox{0.75}{\usym{2613}} & \checkmark &
    {\usym{2613}} & \checkmark \\
  \multicolumn{1}{c|}{} &
    Hassan and Bahsoon \cite{hassan_microservices_2016} &
    2016 &
    \checkmark & \scalebox{0.75}{\usym{2613}} & \scalebox{0.75}{\usym{2613}} & \scalebox{0.75}{\usym{2613}} & \scalebox{0.75}{\usym{2613}} & \checkmark &
    {\usym{2613}} & \checkmark \\
  \multicolumn{1}{c|}{} &
    Lehrig \cite{lehrig_efficiently_2018} &
    2018 &
    \scalebox{0.75}{\usym{2613}} & \scalebox{0.75}{\usym{2613}} & \scalebox{0.75}{\usym{2613}} & \checkmark & \scalebox{0.75}{\usym{2613}} & \checkmark &
    {\usym{2613}} & \checkmark \\
  \multicolumn{1}{c|}{} &
    Medeiros R{\^e}go et al.~\cite{rego2025granularity} &
    2025 &
    \scalebox{0.75}{\usym{2613}} & \scalebox{0.75}{\usym{2613}} & \scalebox{0.75}{\usym{2613}} & \checkmark & \scalebox{0.75}{\usym{2613}} & \checkmark &
    {\usym{2613}} & \checkmark \\
  \multicolumn{1}{c|}{} &
    Fang et al.~\cite{fang2026hgraphscale} &
    2026 &
    \checkmark & \scalebox{0.75}{\usym{2613}} & \scalebox{0.75}{\usym{2613}} & \scalebox{0.75}{\usym{2613}} & \scalebox{0.75}{\usym{2613}} & \checkmark &
    {\usym{2613}} & \checkmark \\ \midrule
  \multicolumn{1}{c|}{\multirow{9}{*}{Maintainability}} &
    Liu et al.~\cite{liu_evaluate_2018} &
    2018 &
    \scalebox{0.75}{\usym{2613}} & \scalebox{0.75}{\usym{2613}} & \scalebox{0.75}{\usym{2613}} & \checkmark & \scalebox{0.75}{\usym{2613}} & \checkmark &
    {\checkmark} & \usym{2613} \\
  \multicolumn{1}{c|}{} &
    Behnamghader et al.~\cite{behnamghader_towards_2017} &
    2017 &
    \scalebox{0.75}{\usym{2613}} & \scalebox{0.75}{\usym{2613}} & \scalebox{0.75}{\usym{2613}} & \checkmark & \scalebox{0.75}{\usym{2613}} & \checkmark &
    {\checkmark} & \usym{2613} \\
  \multicolumn{1}{c|}{} &
    Xie et al.~\cite{xie_statistical_2017} &
    2017 &
    \scalebox{0.75}{\usym{2613}} & \scalebox{0.75}{\usym{2613}} & \scalebox{0.75}{\usym{2613}} & \checkmark & \scalebox{0.75}{\usym{2613}} & \checkmark &
    {\checkmark} & \usym{2613} \\
  \multicolumn{1}{c|}{} &
    Mendonça et al.~\cite{mendonca_generality_2018} &
    2018 &
    \checkmark & \scalebox{0.75}{\usym{2613}} & \scalebox{0.75}{\usym{2613}} & \scalebox{0.75}{\usym{2613}} & \scalebox{0.75}{\usym{2613}} & \checkmark &
    {\usym{2613}} & \checkmark \\
  \multicolumn{1}{c|}{} &
    Baylov and Dimov~\cite{baylov_reference_2018} &
    2018 &
    \checkmark & \scalebox{0.75}{\usym{2613}} & \scalebox{0.75}{\usym{2613}} & \scalebox{0.75}{\usym{2613}} & \scalebox{0.75}{\usym{2613}} & \checkmark &
    {\usym{2613}} & \checkmark \\
  \multicolumn{1}{c|}{} &
    Taibi and Lenarduzzi~\cite{taibi_definition_2018} &
    2018 &
    \scalebox{0.75}{\usym{2613}} & \scalebox{0.75}{\usym{2613}} & \checkmark & \scalebox{0.75}{\usym{2613}} & \scalebox{0.75}{\usym{2613}} & \checkmark &
    {\usym{2613}} & \checkmark \\
  \multicolumn{1}{c|}{} &
    Jaafar et al.~\cite{jaafar_analyzing_2017} &
    2017 &
    \scalebox{0.75}{\usym{2613}} & \scalebox{0.75}{\usym{2613}} & \scalebox{0.75}{\usym{2613}} & \checkmark & \scalebox{0.75}{\usym{2613}} & \checkmark &
    {\usym{2613}} & \checkmark \\
  \multicolumn{1}{c|}{} &
    Schneider et al.~\cite{schneider2025recovery} &
    2025 &
    \scalebox{0.75}{\usym{2613}} & \scalebox{0.75}{\usym{2613}} & \scalebox{0.75}{\usym{2613}} & \checkmark & \scalebox{0.75}{\usym{2613}} & \checkmark &
    {\usym{2613}} & \checkmark \\
  \multicolumn{1}{c|}{} &
    Paudel et al.~\cite{paudel2026technicaldebt} &
    2026 &
    \scalebox{0.75}{\usym{2613}} & \scalebox{0.75}{\usym{2613}} & \scalebox{0.75}{\usym{2613}} & \checkmark & \scalebox{0.75}{\usym{2613}} & \checkmark &
    \checkmark & \checkmark \\ \bottomrule
  \end{tabular}}
  \end{table*}

\subsubsection{Reliability Analysis}
Reliability in service ecosystems refers to the system's ability to consistently provide quality-assured services under both normal and abnormal conditions. This metric reflects the robustness of service design, implementation quality, and fault-handling capabilities throughout the evolution process. We categorize current reliability analysis approaches into two dimensions based on their evaluation focus: recoverability and interoperability.

\textbf{Node Recoverability Analysis.} 
Recoverability refers to a service's ability to resume normal operations following failures or malfunctions through recovery mechanisms such as self-healing and self-adaptation. This analysis typically evaluates individual nodes representing compositions of inseparable services that collaborate to achieve a specific reliability mechanism.

Evaluation methods in this category primarily assess the efficiency and effectiveness of recovery mechanisms. Huang et al.~\cite{huang2015failure} evaluated self-healing capabilities by measuring recovery plan generation time in response to various fault scenarios using a rule-based decision model. While such rule-based approaches may achieve time efficiency, they remain limited by developers' prior knowledge of potential failure modes.

To overcome these limitations, Bashari et al.~\cite{bashari2018self} proposed a knowledge-independent self-healing mechanism that automatically reconfigures product lines to maintain critical functional and non-functional requirements during service failures. Their approach evaluates effectiveness through reconfiguration execution time and self-healing success rates.

Other researchers have focused on improving preliminary recovery components. Sun et al.~\cite{sun2018fault} introduced a fault localization technique for WS-BPEL programs using predicate switching and program slicing, evaluating effectiveness through fault localization success rates and accuracy metrics. Similarly, Asim et al.~\cite{asim2018security} developed a runtime monitoring framework for service nodes, assessing efficiency through the monitoring module's average response time under high loads.

These node-based evaluations primarily focus on the internal reliability mechanisms of isolated service compositions.

Recent studies have further advanced reliability analysis toward network-aware evaluation and robust fault localization. Pham et al.~\cite{pham2024baro} propose BARO, an end-to-end approach integrating anomaly detection and root cause analysis that leverages Multivariate Bayesian Online Change Point Detection to model dependencies within multivariate time-series metrics, demonstrating consistent superiority over existing approaches across three benchmark microservice systems. Zhang et al.~\cite{zhang2025network} propose a network-aware service reliability model that explicitly captures the correlation between network state changes and service reliability, and formulate the microservice placement problem to maximize reliability, reducing service failures by up to 22\% compared to benchmarks.

\textbf{Community Interoperability Analysis.} 
Interoperability refers to a service ecosystem's ability to collaborate and interact effectively with other systems and components. Service nodes with stronger interoperability contribute to higher network reliability by enhancing the ecosystem's resilience against failures.

Community-based reliability analysis evaluates subnetworks or service communities comprehensively. Lu et al.~\cite{lu2020dcem} modeled data services as nodes in bigraphs constructed from structural and message interaction relationships, measuring community reliability by aggregating individual node success rates. Xue et al.~\cite{xue2020integrative, xue2020value} employed information entropy concepts to evaluate node interoperability, proposing a value entropy model to measure collaboration order between service nodes and assess community collaborative efficiency.

Network topology metrics provide another approach for community analysis. Adeleye et al.~\cite{adeleye2018constructing} constructed evolving Web-API communities and evaluated node interoperability through topological properties including degree distribution, network diameter, and clustering coefficients. Building on this approach, Jia et al.~\cite{jia2020research} modeled service influence from a synecological perspective, measuring interaction capabilities through agglomeration degree and effect metrics.

These community-level evaluations either aggregate individual node properties or analyze community topology.

\subsubsection{Usability Analysis}
Usability in a service ecosystem refers to the capability of services to be easily understood, accessed, and utilized by various participants, ensuring a seamless user experience and efficient collaboration. Based on the complex network, existing usability analysis work can be divided into two categories: node-based and community-based.

\textbf{Node-based Usability Analysis} focuses on evaluating and analyzing the usability attributes of individual nodes within the network. 
Initially, Goeb and Lochmann \cite{goeb_software_2011} proposed a software quality model for Service-Oriented Architecture (SOA). This model defines the quality characteristics of SOA systems in five aspects: Service, Service Composition, Business Process, Governance, and Infrastructure. 
To monitor node usability during network runtime, Phipathananunth and Bunyakiati \cite{phipathananunth_synthetic_2018} proposed a runtime monitoring method for microservices architecture. This approach creates a "simulated environment" for service invocation relationships based on the microservices architecture. It then actively calls services within this environment to gather monitoring data, such as latency, failure rate, and throughput. Finally, the collected data is analyzed to detect the architecture's operational status and identify potential issues.
For data acquisition and evaluation, Alkalbani and Hussain \cite{alkalbani_quality_2018} introduced Quality CloudCrowd, a crowdsourced quality assessment platform for SaaS services. This platform involves a large number of minimally trained crowd workers who rate SaaS services and generate evaluation reports.
Considering the availability of Internet of Things (IoT) devices, Stévant et al.~\cite{stevant_optimizing_2018} addressed issues related to mobile device deployment of microservices. They evaluated the robustness of microservice deployment scenarios in typical home networks and mobile smart devices using a simulation platform.

\textbf{Community-based Usability Analysis} primarily assesses the overall usability quality of a service ecosystem to identify better service deployment strategies.
Gribaudo et al.~\cite{gribaudo_performance_2017} combined load testing, network probes, service registries, and queueing models to evaluate system performance under varied loads.
Similarly, Hassan et al.~\cite{hassan_microservices_2016} discussed trade-offs and adaptive methods in microservice design. Their approach involves three levels: strategic, tactical, and operational. It adjusts principles based on strategic-level evaluation results, such as reducing service granularity under high-performance pressure.
Apart from using different testing environments, the relationship between services can be directly analyzed. Lehrig \cite{lehrig_efficiently_2018} explored a method of service quality analysis by modeling architectural knowledge. This method analyzes system architecture documents and source code to identify component and connection relationships. It generates architecture templates containing variables and constraints and calculates service quality indicators for different instances based on the templates.

Recent studies evaluate usability at the system level. Medeiros R{\^e}go et al.~\cite{rego2025granularity} compare different microservice granularities on Amazon EKS. They find that distributing services improves scalability under heavy workloads but increases communication overhead. Fang et al.~\cite{fang2026hgraphscale} model the dependencies among containers, virtual machines, and physical machines for autoscaling, reducing average response time by up to 80.16\% within a fixed cost budget. These results show that usability depends on both service structure and deployment.

In summary, these methods evaluate community-level usability via load testing, architectural modeling, or analyzing service relationships within defined communities.

\subsubsection{Maintainability Analysis}
Maintainability in a service ecosystem refers to the ability of the system to be efficiently updated, adapted, and supported over time, ensuring long-term stability and responsiveness to changing requirements. This is particularly critical in complex, distributed networks where services are interdependent and continuously evolving.

\textbf{Node-based Maintainability Analysis} primarily evaluates the maintainability of individual nodes within a software system by assessing how the software evolves over time.
One intuitive approach to evaluate node maintainability is to use metrics that assess software complexity. Liu et al.~\cite{liu_evaluate_2018} explored the change pattern of cyclomatic complexity during software evolution. They considered cyclomatic complexity as one of the indicators reflecting the difficulty of understanding and testing code. Their study observed a change pattern of increasing-stabilizing-decreasing cyclomatic complexity during software evolution, corresponding to the design-stabilization-refactoring stages of software evolution. 
Another approach, proposed by Behnamghader et al.~\cite{behnamghader_towards_2017}, uses commit history to evaluate node maintainability from an evolutionary perspective. Their method employs commit impact analysis to understand software quality evolution. It establishes correlations between software commits and measured quality metrics, calculates the impact scope of each commit (i.e., the files and code involved), and analyzes the changes within the impact scope to identify the root causes of quality changes.
Furthermore, some individuals evaluate service evolution from an evolutionary perspective by tracking changes in user demands. Xie et al.~\cite{xie_statistical_2017} proposed utilizing statistical analysis methods to predict user demand changes and guide service evolution decisions. Initially, they collected user demand data from various channels, then analyzed the data to generate a statistical model describing the relationship between different demands. Finally, they used the model to predict future demand changes, thus yielding insights into the potential variations in demand.

\textbf{Community-based Maintainability Analysis} assesses the overall maintainability of a system under the network's architecture.
In this context, architectural approaches are commonly used to evaluate system maintainability. Mendonça et al.~\cite{mendonca_generality_2018} introduced an architecture-driven method for building adaptive systems under a microservices architecture. They first design abstract models for microservices and then develop adaptive management mechanisms for each microservice individually. 
Similarly, Baylov and Dimov~\cite{baylov_reference_2018} proposed an adaptive reference architecture for managing system complexity under a microservices architecture. Their reference architecture consists of three levels: management platform, proxy layer, and microservices layer. The management platform collects information from microservices and issues adaptive configurations. The proxy layer translates configurations into rules understandable by individual microservices, and the microservices layer, containing numerous microservices, implements adaptive behaviors based on the rules. 
From an evolutionary perspective, Jaafar et al.~\cite{jaafar_analyzing_2017} proposed an asynchronous change pattern analysis method to analyze software evolution and quality. They pulled all code modification information from the project's history and calculated the change associations between files, determining the causal relationship between different modification commits. Finally, they identified cyclic patterns in the relevant change network, showing the increasing complexity of the project's evolution.

Recent studies assess maintainability using architecture and evolution data. Schneider et al.~\cite{schneider2025recovery} compare nine static architecture-recovery tools. The best tool achieves an F1-score of 0.86, while combining four tools increases it to 0.91. Paudel et al.~\cite{paudel2026technicaldebt} track technical-debt density in one monolith and 78 microservices. They find that technical debt gradually increases across the microservices and varies with service size and team ownership.

In summary, these approaches evaluate maintainability primarily through architectural design principles or historical change analysis.

%% file: section/sec_opportunities.tex
\section{Open Issues and Opportunities}
\label{sec_Open_Issues_and_Opportunities}

Based on our survey through the complex network perspective, we identify three critical open issues and corresponding future opportunities within each stage of service ecosystem evolution:

\begin{itemize}
    \item \textbf{Limited Evolution Perception Models:} Recent studies model hierarchical deployment dependencies and dynamic API correlations~\cite{fang2026hgraphscale,zhong2026popularity}. However, most models are designed for a single task and do not jointly represent service semantics, topology, and runtime behavior~\cite{chang2023dynamic}. Future work should study dynamic network models that combine these sources while controlling data and computational costs, using graph learning methods where appropriate~\cite{scarselli2008graph,wang_heterogeneous_2019}.

    \item \textbf{Limited Support for Multiple Evolution Goals:} Recent studies include network centrality and service-association impact in multi-objective composition~\cite{yang2025service,zhang2025association}. However, their objectives and constraints remain fixed and scenario-specific. Future work should support changing goals and explicit trade-offs among reliability, cost, performance, and maintainability while considering dependencies between services.

    \item \textbf{Lack of Comparable Evaluation Metrics:} Recent studies use response time and cost~\cite{fang2026hgraphscale}, architecture-recovery accuracy~\cite{schneider2025recovery}, and technical-debt density~\cite{paudel2026technicaldebt}. These measures differ in definition, scale, and experimental setting, making direct comparison difficult. Future work should define common metrics for individual services, service groups, and the entire ecosystem, together with reproducible measurement procedures. Measures from ecological network analysis may also provide useful references, but they require clear definitions and validation for service systems~\cite{brundrett1991mycorrhizas,grenni2018ecological,mccormack2021southern}.
\end{itemize}

Addressing these issues and exploring the corresponding opportunities will propel service ecosystem evolution research beyond its current limitations, advancing both theoretical depth and practical application from a complex network perspective.

%% file: section/sec_conclusion.tex
\section{Conclusion}
\label{sec:conclusion}

This paper surveyed service ecosystem evolution from a complex-network perspective and organized the literature through a three-stage framework comprising Perception, Implementation, and Evaluation. The review shows that current research remains limited by insufficiently expressive perception models, fragmented implementation objectives, and inconsistent evaluation metrics. It also highlights opportunities to combine advanced network science, GNNs, and ecobiology-inspired concepts for more predictive and systematic analysis. By connecting structural representation, evolutionary mechanisms, and outcome assessment within one framework, this survey provides a concise foundation for understanding existing work and guiding future research on robust, adaptive service ecosystems.

%% file: main.bbl
\begin{thebibliography}{100}
\providecommand{\url}[1]{#1}
\csname url@samestyle\endcsname
\providecommand{\newblock}{\relax}
\providecommand{\bibinfo}[2]{#2}
\providecommand{\BIBentrySTDinterwordspacing}{\spaceskip=0pt\relax}
\providecommand{\BIBentryALTinterwordstretchfactor}{4}
\providecommand{\BIBentryALTinterwordspacing}{\spaceskip=\fontdimen2\font plus
\BIBentryALTinterwordstretchfactor\fontdimen3\font minus
  \fontdimen4\font\relax}
\providecommand{\BIBforeignlanguage}[2]{{%
\expandafter\ifx\csname l@#1\endcsname\relax
\typeout{** WARNING: IEEEtran.bst: No hyphenation pattern has been}%
\typeout{** loaded for the language `#1'. Using the pattern for}%
\typeout{** the default language instead.}%
\else
\language=\csname l@#1\endcsname
\fi
#2}}
\providecommand{\BIBdecl}{\relax}
\BIBdecl

\bibitem{schroeder2009large}
B.~Schroeder and G.~A. Gibson, ``A large-scale study of failures in
  high-performance computing systems,'' \emph{IEEE transactions on Dependable
  and Secure Computing}, vol.~7, no.~4, pp. 337--350, 2009.

\bibitem{liu2021microhecl}
D.~Liu, C.~He, X.~Peng, F.~Lin, C.~Zhang, S.~Gong, Z.~Li, J.~Ou, and Z.~Wu,
  ``Microhecl: High-efficient root cause localization in large-scale
  microservice systems,'' in \emph{2021 IEEE/ACM 43rd International Conference
  on Software Engineering: Software Engineering in Practice (ICSE-SEIP)}.\hskip
  1em plus 0.5em minus 0.4em\relax IEEE, 2021, pp. 338--347.

\bibitem{deb2013evolutionary}
K.~Deb and H.~Jain, ``An evolutionary many-objective optimization algorithm
  using reference-point-based nondominated sorting approach, part i: solving
  problems with box constraints,'' \emph{IEEE transactions on evolutionary
  computation}, vol.~18, no.~4, pp. 577--601, 2013.

\bibitem{barcelo2016iot}
M.~Barcelo, A.~Correa, J.~Llorca, A.~M. Tulino, J.~L. Vicario, and A.~Morell,
  ``Iot-cloud service optimization in next generation smart environments,''
  \emph{IEEE Journal on Selected Areas in Communications}, vol.~34, no.~12, pp.
  4077--4090, 2016.

\bibitem{gao2017service}
Z.~Gao, Y.~Fan, C.~Wu, W.~Tan, and J.~Zhang, ``Service recommendation from the
  evolution of composition patterns,'' in \emph{2017 IEEE International
  Conference on Services Computing}, 2017, pp. 108--115.

\bibitem{lu2020dcem}
J.~Lu, H.~Zhou, H.~Zhu, Y.~Zhang, Q.~Liang, and G.~Xiao, ``Dcem: A data cell
  evolution model for service composition based on bigraph theory,''
  \emph{Future Generation Computer Systems}, vol. 112, pp. 330--347, 2020.

\bibitem{boley2007digital}
H.~Boley and E.~Chang, ``Digital ecosystems: Principles and semantics,'' in
  \emph{2007 Inaugural IEEE-IES Digital EcoSystems and Technologies
  Conference}, 2007, pp. 398--403.

\bibitem{zhong2014time}
Y.~Zhong, Y.~Fan, K.~Huang, W.~Tan, and J.~Zhang, ``Time-aware service
  recommendation for mashup creation in an evolving service ecosystem,'' in
  \emph{2014 IEEE international conference on web services}, 2014, pp. 25--32.

\bibitem{geary2020guide}
W.~L. Geary, M.~Bode, T.~S. Doherty, E.~A. Fulton, D.~G. Nimmo, A.~I. Tulloch,
  V.~J. Tulloch, and E.~G. Ritchie, ``A guide to ecosystem models and their
  environmental applications,'' \emph{Nature Ecology \& Evolution}, vol.~4,
  no.~11, pp. 1459--1471, 2020.

\bibitem{xue2022research}
X.~Xue, G.~Li, D.~Zhou, Y.~Zhang, L.~Zhang, Y.~Zhao, Z.~Feng, L.~Cui, Z.~Zhou,
  X.~Sun \emph{et~al.}, ``Research roadmap of service ecosystems: A crowd
  intelligence perspective,'' \emph{International Journal of Crowd Science},
  vol.~6, no.~4, pp. 195--222, 2022.

\bibitem{abeywickrama2017survey}
D.~B. Abeywickrama and E.~Ovaska, ``A survey of autonomic computing methods in
  digital service ecosystems,'' \emph{Service Oriented Computing and
  Applications}, vol.~11, pp. 1--31, 2017.

\bibitem{li2011research}
S.~Li and Y.~Fan, ``Research on the service-oriented business ecosystem,'' in
  \emph{2011 3rd International Conference on Advanced Computer Control}, 2011,
  pp. 502--505.

\bibitem{golgeci2022bibliometric}
I.~G{\"o}lgeci, I.~Ali, P.~Ritala, and A.~Arslan, ``A bibliometric review of
  service ecosystems research: current status and future directions,''
  \emph{Journal of Business \& Industrial Marketing}, vol.~37, no.~4, pp.
  841--858, 2022.

\bibitem{erdHos1960evolution}
P.~Erd{\H{o}}s, A.~R{\'e}nyi \emph{et~al.}, ``On the evolution of random
  graphs,'' \emph{Publ. Math. Inst. Hung. Acad. Sci}, vol.~5, no.~1, pp.
  17--60, 1960.

\bibitem{watts1998collective}
D.~J. Watts and S.~H. Strogatz, ``Collective dynamics of
  ‘small-world’networks,'' \emph{nature}, vol. 393, no. 6684, pp. 440--442,
  1998.

\bibitem{barabasi1999emergence}
A.-L. Barab{\'a}si and R.~Albert, ``Emergence of scaling in random networks,''
  \emph{science}, vol. 286, no. 5439, pp. 509--512, 1999.

\bibitem{zhou2020measurement}
Y.~Zhou, Y.~Mi, Y.~Zhu, and L.~Chen, ``Measurement and refactoring for package
  structure based on complex network,'' \emph{Applied Network Science}, vol.~5,
  no.~1, pp. 1--20, 2020.

\bibitem{adeleye2021constructing}
O.~Adeleye, J.~Yu, G.~Wang, and S.~Yongchareon, ``Constructing and evaluating
  evolving web-api networks-a complex network perspective,'' \emph{IEEE
  Transactions on Services Computing}, 2021.

\bibitem{zhou2020evolutionary}
J.~Zhou, L.~Gao, X.~Yao, C.~Zhang, F.~T. Chan, and Y.~Lin, ``Evolutionary
  many-objective assembly of cloud services via angle and adversarial direction
  driven search,'' \emph{Information Sciences}, vol. 513, pp. 143--167.

\bibitem{Papageorgiou2014DecisionSF}
A.~Papageorgiou, A.~Miede, S.~Schulte, D.~Schuller, and R.~Steinmetz,
  ``Decision support for web service adaptation,'' \emph{Pervasive Mob.
  Comput.}, vol.~12, pp. 197--213, 2014.

\bibitem{xue2020value}
X.~Xue, Z.~Chen, S.~Wang, Z.~Feng, Y.~Duan, and Z.~Zhou, ``Value entropy: A
  systematic evaluation model of service ecosystem evolution,'' \emph{IEEE
  Transactions on Services Computing}, vol.~15, no.~4, pp. 1760--1773, 2020.

\bibitem{xue2020integrative}
X.~Xue, S.~Chen, B.~Li, Z.~Chen, and S.~Wang, ``An integrative
  multi-dimensional evaluation of service ecosystem,'' in \emph{2020 IEEE
  International Conference on Web Services}, 2020, pp. 265--272.

\bibitem{cao2019qos}
B.~Cao, J.~Liu, Y.~Wen, H.~Li, Q.~Xiao, and J.~Chen, ``Qos-aware service
  recommendation based on relational topic model and factorization machines for
  iot mashup applications,'' \emph{Journal of parallel and distributed
  computing}, vol. 132, pp. 177--189, 2019.

\bibitem{jiang2019cloud}
Y.~Jiang, D.~Tao, Y.~Liu, J.~Sun, and H.~Ling, ``Cloud service recommendation
  based on unstructured textual information,'' \emph{Future Generation Computer
  Systems}, vol.~97, pp. 387--396, 2019.

\bibitem{yang2021web}
D.~Yang and D.~He, ``Web service clustering method based on word vector and
  biterm topic model,'' in \emph{2021 IEEE 6th International Conference on
  Cloud Computing and Big Data Analytics}, 2021, pp. 299--304.

\bibitem{shen2022picf}
J.~Shen, W.~Huang, and Q.~Hu, ``Picf-lda: a topic enhanced lda with probability
  incremental correction factor for web api service clustering,'' \emph{Journal
  of Cloud Computing}, vol.~11, no.~1, pp. 1--13, 2022.

\bibitem{zhang2020learning}
J.~Zhang, Y.~Fan, J.~Zhang, and B.~Bai, ``Learning to build accurate service
  representations and visualization,'' \emph{IEEE Transactions on Services
  Computing}, vol.~15, no.~3, pp. 1551--1563, 2020.

\bibitem{chen2013constructing}
W.~Chen, I.~Paik, and P.~C. Hung, ``Constructing a global social service
  network for better quality of web service discovery,'' \emph{IEEE
  transactions on services computing}, vol.~8, no.~2, pp. 284--298, 2013.

\bibitem{lyu2014three}
S.~Lyu, J.~Liu, M.~Tang, G.~Kang, B.~Cao, and Y.~Duan, ``Three-level views of
  the web service network: an empirical study based on programmableweb,'' in
  \emph{2014 ieee international congress on big data}, 2014, pp. 374--381.

\bibitem{liu2023data}
M.~Liu, Z.~Tu, X.~Xu, Z.~Wang, and Y.~Wang, ``A data-driven approach for
  constructing multilayer network-based service ecosystem models,''
  \emph{Software and Systems Modeling}, vol.~22, no.~3, pp. 919--939, 2023.

\bibitem{adeleye2019fitness}
O.~Adeleye, J.~Yu, S.~Yongchareon, Q.~Z. Sheng, and L.~H. Yang, ``A
  fitness-based evolving network for web-apis discovery,'' in \emph{Proceedings
  of the australasian computer science week multiconference}, 2019, pp. 1--10.

\bibitem{wang2018mashup}
X.~Wang, H.~Wu, and C.-H. Hsu, ``Mashup-oriented api recommendation via random
  walk on knowledge graph,'' \emph{IEEE Access}, vol.~7, pp. 7651--7662, 2018.

\bibitem{cao2024prkg}
B.~Cao, M.~Peng, Z.~Xie, J.~Liu, H.~Ye, B.~Li, and K.~K. Fletcher, ``Prkg:
  Pre-training representation and knowledge-graph-enhanced web service
  recommendation for mashup creation,'' \emph{IEEE Transactions on Network and
  Service Management}, vol.~21, no.~2, pp. 1737--1749, 2024.

\bibitem{tang2024light}
M.~Tang, J.~Mai, F.~Xie, and Z.~Zheng, ``Light heterogeneous hypergraph
  contrastive learning based service recommendation for mashup creation,''
  \emph{IEEE Transactions on Services Computing}, vol.~17, no.~6, pp.
  3844--3856, 2024.

\bibitem{gao2019discovery}
Z.~Gao, Y.~Fan, X.~Li, L.~Gu, C.~Wu, and J.~Zhang, ``Discovery and analysis
  about the evolution of service composition patterns,'' \emph{Journal of Web
  Engineering}, pp. 579--626, 2019.

\bibitem{cao2017integrated}
B.~Cao, X.~F. Liu, M.~M. Rahman, B.~Li, J.~Liu, and M.~Tang, ``Integrated
  content and network-based service clustering and web apis recommendation for
  mashup development,'' \emph{IEEE Transactions on Services Computing},
  vol.~13, no.~1, pp. 99--113, 2017.

\bibitem{huang2014recommendation}
K.~Huang, Y.~Fan, and W.~Tan, ``Recommendation in an evolving service ecosystem
  based on network prediction,'' \emph{IEEE Transactions on Automation Science
  and Engineering}, vol.~11, no.~3, pp. 906--920, 2014.

\bibitem{adeleye2020complex}
O.~Adeleye, J.~Yu, S.~Yongchareon~Yongchareon, Y.~Han, and Q.~Sheng, ``Complex
  network-based web service for web-api discovery,'' in \emph{Proceedings of
  the australasian computer science week multiconference}, 2020, pp. 1--10.

\bibitem{liu2021dysr}
M.~Liu, Z.~Tu, X.~Xu, and Z.~Wang, ``Dysr: A dynamic representation learning
  and aligning based model for service bundle recommendation,'' \emph{arXiv
  preprint arXiv:2108.03360}, 2021.

\bibitem{he2025graph}
Y.~He, Z.~Feng, H.~Chen, J.~Li, Q.~Gao, H.~Yi, and H.~Zhang, ``Graph neural
  networks for incremental service recommendation with dynamic interest
  alignment,'' in \emph{2025 IEEE International Conference on Web Services
  (ICWS)}.\hskip 1em plus 0.5em minus 0.4em\relax IEEE, 2025, pp. 1--10.

\bibitem{zhou2024platform}
X.~Zhou, J.~Xiao, X.~Xue, S.~Chen, and Z.~Feng, ``A platform ecosystem
  evolution model with service dynamic supply and matching,'' \emph{IEEE
  Transactions on Computational Social Systems}, vol.~11, no.~3, pp.
  3504--3515, 2024.

\bibitem{zhong2026popularity}
W.~Zhong, D.~Zhai, A.~K. Fakhrabadi, H.~Attar, Y.~Yan, R.~Jiang, and S.~Wang,
  ``Exploring and mitigating the impact of popularity bias for dynamic {API}
  composition recommendations,'' \emph{Tsinghua Science and Technology},
  vol.~31, no.~2, pp. 1233--1247, 2026.

\bibitem{zhang2019mining}
N.~Zhang, J.~Wang, K.~He, Z.~Li, and Y.~Huang, ``Mining and clustering service
  goals for restful service discovery,'' \emph{Knowledge and Information
  Systems}, vol.~58, pp. 669--700, 2019.

\bibitem{bianconi2001bose}
G.~Bianconi and A.-L. Barab{\'a}si, ``Bose-einstein condensation in complex
  networks,'' \emph{Physical review letters}, vol.~86, no.~24, p. 5632, 2001.

\bibitem{bianconi2001competition}
------, ``Competition and multiscaling in evolving networks,''
  \emph{Europhysics letters}, vol.~54, no.~4, p. 436, 2001.

\bibitem{lda2003}
\BIBentryALTinterwordspacing
D.~M. Blei, A.~Y. Ng, and M.~I. Jordan, ``Latent dirichlet allocation,''
  \emph{J. Mach. Learn. Res.}, vol.~3, pp. 993--1022, 2003. [Online].
  Available: \url{https://jmlr.org/papers/v3/blei03a.html}
\BIBentrySTDinterwordspacing

\bibitem{maamar2011using}
Z.~Maamar, N.~Faci, L.~Wives, Y.~Badr, P.~Santos, and J.~P.~M. De~Oliveira,
  ``Using social networks for web services discovery,'' \emph{IEEE internet
  computing}, vol.~15, no.~4, pp. 48--54, 2011.

\bibitem{6928899}
S.~Wang, W.~A. Higashino, M.~Hayes, and M.~A.~M. Capretz, ``Service evolution
  patterns,'' in \emph{2014 IEEE International Conference on Web Services},
  2014, pp. 201--208.

\bibitem{liu2020novel}
M.~Liu, Z.~Tu, J.~Wang, and Z.~Wang, ``A novel multi-layer network model for
  service ecosystems,'' in \emph{2020 international conference on service
  science}, 2020, pp. 23--30.

\bibitem{zhou2022sle2}
D.~Zhou, X.~Xue, and Z.~Zhou, ``Sle2: The improved social learning evolution
  model of cloud manufacturing service ecosystem,'' \emph{IEEE Transactions on
  Industrial Informatics}, vol.~18, no.~12, pp. 9017--9026, 2022.

\bibitem{fallatah2014social}
H.~Fallatah, J.~Bentahar, and E.~K. Asl, ``Social network-based framework for
  web services discovery,'' in \emph{2014 international conference on future
  internet of things and cloud}, 2014, pp. 159--166.

\bibitem{tao2024diagnosing}
L.~Tao, X.~Lu, S.~Zhang, J.~Luan, Y.~Li, M.~Li, Z.~Li, Q.~Yu, H.~Xie, R.~Xu
  \emph{et~al.}, ``Diagnosing performance issues for large-scale microservice
  systems with heterogeneous graph,'' \emph{IEEE Transactions on Services
  Computing}, vol.~17, no.~5, pp. 2223--2235, 2024.

\bibitem{li2025tracedae}
J.~Li, S.~Ying, T.~Li, and X.~Tian, ``Tracedae: Trace-based anomaly detection
  in micro-service systems via dual autoencoder,'' \emph{IEEE Transactions on
  Network and Service Management}, 2025.

\bibitem{cerny2025analyzing}
T.~Cerny, G.~Goulis, S.~Perry, M.~Edmonds, A.~S. Abdelfattah, M.~Esposito,
  A.~Bakhtin, V.~Lenarduzzi, and D.~Taibi, ``Analyzing evolution of
  microservice-based systems,'' in \emph{Proceedings of the 33rd ACM
  International Conference on the Foundations of Software Engineering}, 2025,
  pp. 1030--1034.

\bibitem{palla2007quantifying}
G.~Palla, A.-L. Barab{\'a}si, and T.~Vicsek, ``Quantifying social group
  evolution,'' \emph{Nature}, vol. 446, no. 7136, pp. 664--667, 2007.

\bibitem{brodka2013ged}
P.~Br{\'o}dka, S.~Saganowski, and P.~Kazienko, ``Ged: the method for group
  evolution discovery in social networks,'' \emph{Social Network Analysis and
  Mining}, vol.~3, pp. 1--14, 2013.

\bibitem{liu2022community}
M.~Liu, Z.~Tu, H.~Xu, X.~Xu, and Z.~Wang, ``Community-based service ecosystem
  evolution analysis,'' \emph{Service Oriented Computing and Applications},
  vol.~16, no.~2, pp. 97--110, 2022.

\bibitem{li2018role}
C.~Li, B.~Q. Feng, Y.~M. Li, S.~L. Hu, N.~Yang, and C.~J. Tang, ``Role-based
  structural evolution and prediction in dynamic networks,'' \emph{Journal of
  Software}, vol.~28, no.~3, pp. 663--675, 2018.

\bibitem{wang2016prediction}
H.~Wang, M.~Kessentini, and A.~Ouni, ``Prediction of web services evolution,''
  in \emph{Service-Oriented Computing: 14th International Conference}, 2016,
  pp. 282--297.

\bibitem{zhu2022software}
X.~Zhu, N.~Li, and Y.~Wang, ``Software change-proneness prediction based on
  deep learning,'' \emph{Journal of Software: Evolution and Process}, vol.~34,
  no.~4, p. e2434, 2022.

\bibitem{wang2021self}
X.~Wang, Z.~Feng, K.~Huang, and S.~Chen, ``Self-adaptation and distributed
  knowledge-based service ecosystem evolution,'' \emph{Concurrency and
  Computation: Practice and Experience}, vol.~33, no.~24, p. e6469, 2021.

\bibitem{chaturvedi2020service}
A.~Chaturvedi, A.~Tiwari, D.~Binkley, and S.~Chaturvedi, ``Service evolution
  analytics: Change and evolution mining of a distributed system,'' \emph{IEEE
  Transactions on Engineering Management}, vol.~68, no.~1, pp. 137--148, 2020.

\bibitem{tan2021method}
W.~Tan, L.~Huang, M.~Y. Kataev, Y.~Sun, L.~Zhao, H.~Zhu, K.~Guo, and N.~Xie,
  ``Method towards reconstructing collaborative business processes with cloud
  services using evolutionary deep q-learning,'' \emph{Journal of Industrial
  Information Integration}, vol.~21, p. 100189, 2021.

\bibitem{huang2025qos}
H.~Huang, Q.~Wang, G.~Min, M.~Wang, and D.~O. Wu, ``Qos prediction for
  component services in 5 g via graph-based deep reinforcement learning,''
  \emph{IEEE Transactions on Mobile Computing}, 2025.

\bibitem{stocker2021code}
M.~Stocker and O.~Zimmermann, ``From code refactoring to api refactoring: Agile
  service design and evolution,'' in \emph{Service-Oriented Computing: 15th
  Symposium and Summer School}, 2021, pp. 174--193.

\bibitem{briscoe2006digital}
G.~Briscoe and P.~De~Wilde, ``Digital ecosystems: evolving service-orientated
  architectures,'' in \emph{Proceedings of the 1st international conference on
  Bio inspired models of network, information and computing systems}, 2006, pp.
  17--es.

\bibitem{Ramrez2017EvolutionaryCO}
A.~Ram{\'i}rez, J.~A. Parejo, J.~R. Romero, S.~Segura, and A.~Ruiz-Cort{\'e}s,
  ``Evolutionary composition of qos-aware web services: A many-objective
  perspective,'' \emph{Expert Syst. Appl.}, vol.~72, pp. 357--370, 2017.

\bibitem{camilli2024integrated}
M.~Camilli, F.~Luccioletti, R.~Mirandola, and P.~Scandurra, ``Integrated
  qos-and vulnerability-driven self-adaptation for microservices
  applications,'' in \emph{International Conference on Service-Oriented
  Computing}.\hskip 1em plus 0.5em minus 0.4em\relax Springer, 2024, pp.
  55--71.

\bibitem{aubonnet2015management}
T.~Aubonnet, L.~Henrio, S.~Kessal, O.~Kulankhina, F.~Lemoine, E.~Madelaine,
  C.~Ruz, and N.~Simoni, ``Management of service compositionbased on
  self-controlled components,'' \emph{Journal of Internet Services and
  Applications}, vol.~6, no.~1, pp. 1--17, 2015.

\bibitem{Lv2018EfficientDE}
C.~Lv, W.~Jiang, S.~Hu, J.~Wang, G.~Lu, and Z.~Liu, ``Efficient dynamic
  evolution of service composition,'' \emph{IEEE Transactions on Services
  Computing}, vol.~11, pp. 630--643, 2018.

\bibitem{Wang2020IntegratingRN}
H.~Wang, J.~Li, Q.~Yu, T.~Hong, J.~Yan, and W.~Zhao, ``Integrating recurrent
  neural networks and reinforcement learning for dynamic service composition,''
  \emph{Future Gener. Comput. Syst.}, vol. 107, pp. 551--563, 2020.

\bibitem{hochreiter1997long}
S.~Hochreiter and J.~Schmidhuber, ``Long short-term memory,'' \emph{Neural
  computation}, vol.~9, no.~8, pp. 1735--1780, 1997.

\bibitem{Jatoth2019OptimalFA}
C.~Jatoth, G.~R. Gangadharan, and R.~Buyya, ``Optimal fitness aware cloud
  service composition using an adaptive genotypes evolution based genetic
  algorithm,'' \emph{Future Gener. Comput. Syst.}, vol.~94, pp. 185--198, 2019.

\bibitem{Cao2012}
G.~Cao, Q.~Tan, and H.~Wu, ``Research on dynamic web service behavior
  adaptation,'' \emph{Computer Science}, vol.~39, 2012.

\bibitem{Kuang2009}
K.~Li, ``Research on web services discovery and adaptation based on interface
  and behavioral semantics,'' 2009.

\bibitem{Benatallah2005DevelopingAF}
B.~Benatallah, F.~Casati, D.~Grigori, H.~R.~M. Nezhad, and F.~Toumani,
  ``Developing adapters for web services integration,'' in \emph{International
  Conference on Advanced Information Systems Engineering}, 2005.

\bibitem{Yellin1997ProtocolSA}
D.~M. Yellin and R.~E. Strom, ``Protocol specifications and component
  adaptors,'' \emph{ACM Trans. Program. Lang. Syst.}, vol.~19, pp. 292--333,
  1997.

\bibitem{cao2013approach}
G.~Cao, Q.~Tan, and H.~Wu, ``An approach to web service adaptation based on
  semantic service flow nets,'' \emph{Journal of Chinese Computer Systems},
  vol.~34, no.~12, pp. 2696--2701, 2013.

\bibitem{SpringAOP}
Spring, ``Spring framework,'' \url{https://spring.io/}.

\bibitem{zhong2024refactoring}
C.~Zhong, S.~Li, H.~Zhang, H.~Huang, L.~Yang, and Y.~Cai, ``Refactoring
  microservices to microservices in support of evolutionary design,''
  \emph{IEEE transactions on software engineering}, vol.~51, no.~2, pp.
  484--502, 2024.

\bibitem{xu2025ggrme}
Y.~Xu, Y.~Li, S.~Wu, L.~Li, X.~Zhu, M.~Xi, and J.~Yin, ``Ggrme: A ggnn-based
  graph reconstruction method for microservice extraction,'' in \emph{2025 IEEE
  International Conference on Web Services (ICWS)}.\hskip 1em plus 0.5em minus
  0.4em\relax IEEE, 2025, pp. 976--982.

\bibitem{xing2025saber}
Y.~Xing, Y.~Lv, X.~Zeng, B.~Zhao, K.~Zhang, H.~Sun, W.~Yang, and Z.~Tu,
  ``Saber: A mape-k-based self-adaptive framework for microservice bad smell
  refactoring,'' in \emph{2025 IEEE International Conference on Web Services
  (ICWS)}.\hskip 1em plus 0.5em minus 0.4em\relax IEEE, 2025, pp. 673--684.

\bibitem{Taleb2013AnAM}
T.~Taleb and A.~Ksentini, ``An analytical model for follow me cloud,''
  \emph{2013 IEEE Global Communications Conference}, pp. 1291--1296, 2013.

\bibitem{Machen2016MigratingRA}
A.~Machen, S.~Wang, K.~K. Leung, B.~Ko, and T.~Salonidis, ``Migrating running
  applications across mobile edge clouds: poster,'' \emph{Proceedings of the
  22nd Annual International Conference on Mobile Computing and Networking},
  2016.

\bibitem{Wang2011ExploitingMP}
J.~Wang, ``Exploiting mobility prediction for dependable service composition in
  wireless mobile ad hoc networks,'' \emph{IEEE Transactions on Services
  Computing}, vol.~4, pp. 44--55, 2011.

\bibitem{Deng2016TowardRR}
S.~Deng, L.~Huang, Y.~Li, H.~Zhou, Z.~Wu, X.~Cao, M.~Y. Kataev, and L.~X. Li,
  ``Toward risk reduction for mobile service composition,'' \emph{IEEE
  Transactions on Cybernetics}, vol.~46, pp. 1807--1816, 2016.

\bibitem{Chen2014ADS}
N.~Chen and S.~Clarke, ``A dynamic service composition model for adaptive
  systems in mobile computing environments,'' in \emph{International Conference
  on Service Oriented Computing}, 2014.

\bibitem{liujiwei2016}
J.~Liu and X.~Mao, ``Research on dynamic evolution mechanisms of software
  runtime variability,'' \emph{CHINESE JOURNAL OF COMPUTERS}, vol.~39, no.~11,
  pp. 2216--2235, 2016.

\bibitem{yang2025service}
T.~Yang, F.~Jiang, and J.~Su, ``A service composition optimization approach for
  cloud manufacturing based on the dual-layer coupled network and ipso
  algorithm,'' \emph{Computers \& Industrial Engineering}, p. 111529, 2025.

\bibitem{zhang2025association}
C.~Zhang, L.~Wang, and K.~He, ``Cloud service composition optimization based on
  service association impact and improved {NSGA-II} algorithm,''
  \emph{Scientific Reports}, vol.~15, no.~1, p. 26001, 2025.

\bibitem{chenzhaojie2021}
Z.~Zhao, J.~Wang, and X.~Xue, ``Entropy model-based analysis method of service
  ecosystem evolution,'' \emph{Application Research of Computers}, vol.~38,
  no.~1, 2021.

\bibitem{Ha2015AdaptiveVH}
K.~Ha, Y.~Abe, Z.~Chen, W.~Hu, B.~Amos, P.~Pillai, and M.~Satyanarayanan,
  ``Adaptive vm handoff across cloudlets,'' 2015.

\bibitem{huang2015failure}
H.~Huang, X.~Chen, and Z.~Wang, ``Failure recovery in distributed model
  composition with intelligent assistance,'' \emph{Information Systems
  Frontiers}, vol.~17, pp. 673--689, 2015.

\bibitem{bashari2018self}
M.~Bashari, E.~Bagheri, and W.~Du, ``Self-adaptation of service compositions
  through product line reconfiguration,'' \emph{Journal of Systems and
  Software}, vol. 144, pp. 84--105, 2018.

\bibitem{margaris2016improving}
D.~Margaris, C.~Vassilakis, and P.~Georgiadis, ``Improving qos delivered by
  ws-bpel scenario adaptation through service execution parallelization,'' in
  \emph{Proceedings of the 31st Annual ACM Symposium on Applied Computing},
  2016, pp. 1590--1596.

\bibitem{jia2020research}
Z.~Jia, S.~Huang, and Y.~Fan, ``Research on the synecological model and dynamic
  evolution mechanism of service internet,'' in \emph{2020 IEEE International
  Conference on Services Computing}, 2020, pp. 12--19.

\bibitem{adeleye2018constructing}
O.~Adeleye, J.~Yu, S.~Yongchareon, and Y.~Han, ``Constructing and evaluating an
  evolving web-api network for service discovery,'' in \emph{Service-Oriented
  Computing: 16th International Conference}, 2018, pp. 603--617.

\bibitem{alferez2017achieving}
G.~H. Alf{\'e}rez and V.~Pelechano, ``Achieving autonomic web service
  compositions with models at runtime,'' \emph{Computers \& Electrical
  Engineering}, vol.~63, pp. 332--352, 2017.

\bibitem{subramanian2008enhancement}
S.~Subramanian, P.~Thiran, N.~C. Narendra, G.~K. Mostefaoui, and Z.~Maamar,
  ``On the enhancement of bpel engines for self-healing composite web
  services,'' in \emph{2008 International Symposium on Applications and the
  Internet}.\hskip 1em plus 0.5em minus 0.4em\relax IEEE, 2008, pp. 33--39.

\bibitem{asim2018security}
M.~Asim, A.~Yautsiukhin, A.~D. Brucker, T.~Baker, Q.~Shi, and B.~Lempereur,
  ``Security policy monitoring of bpmn-based service compositions,''
  \emph{Journal of Software: Evolution and Process}, vol.~30, no.~9, p. e1944,
  2018.

\bibitem{alkalbani_quality_2018}
A.~M. Alkalbani and F.~K. Hussain, ``Quality {CloudCrowd}: A crowdsourcing
  platform for {QoS} assessment of saas services,'' in \emph{Advances on P2P,
  Parallel, Grid, Cloud and Internet Computing: Proceedings of the 12th
  International Conference on P2P, Parallel, Grid, Cloud and Internet
  Computing}, pp. 235--240.

\bibitem{taibi_definition_2018}
D.~Taibi and V.~Lenarduzzi, ``On the definition of microservice bad smells,''
  vol.~35, no.~3, pp. 56--62.

\bibitem{jaafar_analyzing_2017}
F.~Jaafar, A.~Lozano, Y.-G. Guéhéneuc, and K.~Mens, ``Analyzing software
  evolution and quality by extracting asynchrony change patterns,'' vol. 131,
  pp. 311--322.

\bibitem{sun2018fault}
C.-a. Sun, Y.~Ran, C.~Zheng, H.~Liu, D.~Towey, and X.~Zhang, ``Fault
  localisation for ws-bpel programs based on predicate switching and program
  slicing,'' \emph{Journal of Systems and Software}, vol. 135, pp. 191--204,
  2018.

\bibitem{pham2024baro}
L.~Pham, H.~Ha, and H.~Zhang, ``Baro: Robust root cause analysis for
  microservices via multivariate bayesian online change point detection,''
  \emph{Proceedings of the ACM on Software Engineering}, vol.~1, no. FSE, pp.
  2214--2237, 2024.

\bibitem{zhang2025network}
F.~Zhang, Y.~Chen, H.~Lu, and Y.~Huang, ``Network-aware reliability modeling
  and optimization for microservice placement,'' \emph{IEEE Transactions on
  Network and Service Management}, 2025.

\bibitem{goeb_software_2011}
A.~Goeb and K.~Lochmann, ``A software quality model for {SOA},'' in
  \emph{Proceedings of the 8th International Workshop on Software Quality}, pp.
  18--25.

\bibitem{phipathananunth_synthetic_2018}
C.~Phipathananunth and P.~Bunyakiati, ``Synthetic runtime monitoring of
  microservices software architecture,'' in \emph{2018 {IEEE} 42nd Annual
  Computer Software and Applications Conference}, vol.~02, pp. 448--453.

\bibitem{stevant_optimizing_2018}
B.~Stévant, J.-L. Pazat, and A.~Blanc, ``Optimizing the performance of a
  microservice-based application deployed on user-provided devices,'' in
  \emph{2018 17th International Symposium on Parallel and Distributed
  Computing}, pp. 133--140.

\bibitem{gribaudo_performance_2017}
M.~Gribaudo, M.~Iacono, D.~Manini, and {others}, ``Performance evaluation of
  massively distributed microservices based applications,'' in
  \emph{Proceedings-31st European Conference on Modelling and Simulation}, pp.
  598--604.

\bibitem{avritzer_quantitative_2018}
A.~Avritzer, V.~Ferme, A.~Janes, B.~Russo, H.~Schulz, and A.~van Hoorn, ``A
  quantitative approach for the assessment of microservice architecture
  deployment alternatives by automated performance testing,'' in \emph{Software
  Architecture: 12th European Conference on Software Architecture}, pp.
  159--174.

\bibitem{hassan_microservices_2016}
S.~Hassan and R.~Bahsoon, ``Microservices and their design trade-offs: A
  self-adaptive roadmap,'' in \emph{2016 {IEEE} International Conference on
  Services Computing}, pp. 813--818.

\bibitem{lehrig_efficiently_2018}
S.~M. Lehrig, \emph{Efficiently conducting quality-of-service analyses by
  templating architectural knowledge}, vol.~25.

\bibitem{rego2025granularity}
C.~Medeiros~R{\^e}go, R.~C. Mendon{\c{c}}a~Filho, and N.~C. Mendon{\c{c}}a,
  ``Performance and resilience impact of microservice granularity: An empirical
  evaluation using {Service Weaver} and {Amazon EKS},'' \emph{International
  Journal of Network Management}, vol.~35, no.~4, p. e70019, 2025.

\bibitem{fang2026hgraphscale}
Z.~Fang, H.~Ma, G.~Chen, and R.~Buyya, ``{HGraphScale}: Hierarchical graph
  learning for autoscaling microservice applications in container-based cloud
  computing,'' \emph{IEEE Transactions on Services Computing}, vol.~19, no.~1,
  pp. 410--422, 2026.

\bibitem{liu_evaluate_2018}
H.~Liu, X.~Gong, L.~Liao, and B.~Li, ``Evaluate how cyclomatic complexity
  changes in the context of software evolution,'' in \emph{2018 {IEEE} 42nd
  Annual Computer Software and Applications Conference}, vol.~2, pp. 756--761.

\bibitem{behnamghader_towards_2017}
P.~Behnamghader, R.~Alfayez, K.~Srisopha, and B.~Boehm, ``Towards better
  understanding of software quality evolution through commit-impact analysis,''
  in \emph{2017 {IEEE} International conference on software quality,
  reliability and security}, pp. 251--262.

\bibitem{xie_statistical_2017}
H.~Xie, J.~Yang, C.~K. Chang, and L.~Liu, ``A statistical analysis approach to
  predict user's changing requirements for software service evolution,'' vol.
  132, pp. 147--164.

\bibitem{mendonca_generality_2018}
N.~C. Mendonça, D.~Garlan, B.~Schmerl, and J.~Cámara, ``Generality vs.
  reusability in architecture-based self-adaptation: The case for self-adaptive
  microservices,'' in \emph{Proceedings of the 12th European Conference on
  Software Architecture: Companion Proceedings}, pp. 1--6.

\bibitem{baylov_reference_2018}
K.~Baylov and A.~Dimov, ``Reference architecture for self-adaptive microservice
  systems,'' pp. 297--303.

\bibitem{schneider2025recovery}
S.~Schneider, A.~Bakhtin, X.~Li, J.~Soldani, A.~Brogi, T.~Cerny,
  R.~Scandariato, and D.~Taibi, ``Comparison of static analysis architecture
  recovery tools for microservice applications,'' \emph{Empirical Software
  Engineering}, vol.~30, no.~5, p. 128, 2025.

\bibitem{paudel2026technicaldebt}
B.~Paudel, J.~Gonzalez-Huerta, and E.~Zabardast, ``Exploring the evolution of
  technical debt in monolithic and hybrid microservice architecture: An
  industrial case study,'' \emph{Journal of Systems and Software}, vol. 237, p.
  112831, 2026.

\bibitem{chang2023dynamic}
F.~Chang, G.~Zhou, Q.~Huang, K.~Ding, W.~Cheng, J.~Hui, Y.~Zhi, and C.~Zhang,
  ``A dynamic multi-layer maintenance service network evolution and
  decision-making method for service-oriented complex equipment,''
  \emph{Computers \& Industrial Engineering}, vol. 181, p. 109319, 2023.

\bibitem{scarselli2008graph}
F.~Scarselli, M.~Gori, A.~C. Tsoi, M.~Hagenbuchner, and G.~Monfardini, ``The
  graph neural network model,'' \emph{IEEE Transactions on Neural Networks},
  vol.~20, no.~1, pp. 61--80, 2008.

\bibitem{wang_heterogeneous_2019}
X.~Wang, H.~Ji, C.~Shi, B.~Wang, Y.~Ye, P.~Cui, and P.~S. Yu, ``Heterogeneous
  {Graph} {Attention} {Network},'' 2019, pp. 2022--2032.

\bibitem{brundrett1991mycorrhizas}
M.~Brundrett, ``Mycorrhizas in natural ecosystems,'' in \emph{Advances in
  ecological research}.\hskip 1em plus 0.5em minus 0.4em\relax Elsevier, 1991,
  vol.~21, pp. 171--313.

\bibitem{grenni2018ecological}
P.~Grenni, V.~Ancona, and A.~B. Caracciolo, ``Ecological effects of antibiotics
  on natural ecosystems: A review,'' \emph{Microchemical Journal}, vol. 136,
  pp. 25--39, 2018.

\bibitem{mccormack2021southern}
S.~A. McCormack, J.~Melbourne-Thomas, R.~Trebilco, G.~Griffith, S.~L. Hill,
  C.~Hoover, N.~M. Johnston, T.~I. Marina, E.~J. Murphy, E.~A. Pakhomov
  \emph{et~al.}, ``Southern ocean food web modelling: Progress, prognoses, and
  future priorities for research and policy makers,'' \emph{Frontiers in
  Ecology and Evolution}, p. 626, 2021.

\end{thebibliography}
